\documentclass[twocolumn]{emulateapj}

\usepackage{natbib}

\lefthead{Law et al. 2012}
\righthead{Relation of $z\sim2$ Morphologies to Gas Kinematics}

\slugcomment{DRAFT: \today}

\begin{document}

\newcommand{\msun}{\ensuremath{\rm M_\odot}}
\newcommand{\msunyr}{\ensuremath{\rm M_{\odot}\;{\rm yr}^{-1}}}
\newcommand{\Ha}{\ensuremath{\rm H\alpha}}
\newcommand{\Hb}{\ensuremath{\rm H\beta}}
\newcommand{\lya}{\ensuremath{\rm Ly\alpha}}
\newcommand{\Ntwo}{[\ion{N}{2}]}
\newcommand{\kms}{\textrm{km~s}\ensuremath{^{-1}\,}}
\newcommand{\ztwo}{\ensuremath{z\sim2}}
\newcommand{\zthree}{\ensuremath{z\sim3}}
\newcommand{\feh}{\textrm{[Fe/H]}}
\newcommand{\afeh}{\textrm{[$\alpha$/Fe]}}
\newcommand{\nifeh}{\textrm{[Ni/Fe]}}
\newcommand{\othree}{\textrm{[O\,{\sc iii}]}}
\newcommand{\otwo}{\textrm{[O\,{\sc ii}]}}
\newcommand{\ntwo}{\textrm{[N\,{\sc ii}]}}

\newcommand{\sitwo}{\textrm{Si\,{\sc ii}}}
\newcommand{\oone}{\textrm{O\,{\sc i}}}
\newcommand{\ctwo}{\textrm{C\,{\sc ii}}}
\newcommand{\sifour}{\textrm{Si\,{\sc iv}}}
\newcommand{\cfour}{\textrm{C\,{\sc iv}}}
\newcommand{\fetwo}{\textrm{Fe\,{\sc ii}}}
\newcommand{\altwo}{\textrm{Al\,{\sc ii}}}
\newcommand{\hetwo}{\textrm{He\,{\sc ii}}}
\newcommand{\ciii}{\textrm{C\,{\sc iii}]}}

\newcommand{\dvis}{\ensuremath{\Delta v_{\rm IS}}}
\newcommand{\hst}{{\it HST}-WFC3}

\shortauthors{Law et al.}
\shorttitle{Relation of $z\sim2$ Morphologies to Gas Kinematics}

\title{An HST/WFC3-IR Morphological Survey of Galaxies at $z = 1.5-3.6$: II.  The Relation between Morphology and Gas-Phase Kinematics}\thanks{Based 
in part on data obtained at the W. M. Keck Observatory, which is operated as a scientific partnership among the California Institute of Technology, the University of 
California, and NASA, and was made possible by the generous financial support of the W. M. Keck Foundation.}

\author{David R. Law\altaffilmark{1}, Charles C. Steidel\altaffilmark{2}, Alice E. Shapley\altaffilmark{3}, Sarah R. Nagy\altaffilmark{3}, Naveen A. Reddy\altaffilmark{4}, \& Dawn K. Erb\altaffilmark{5}}

\altaffiltext{1}{Dunlap Fellow, Dunlap Institute for Astronomy \& Astrophysics, University of Toronto, 50 St. George Street, Toronto M5S 3H4, Ontario, Canada; drlaw@di.utoronto.ca}
\altaffiltext{2}{California Institute of Technology, MS 249-17, Pasadena, CA 91125; ccs@astro.caltech.edu}
\altaffiltext{3}{Department of Physics and Astronomy, University of California, Los Angeles, CA 90095;
aes@astro.ucla.edu, snagy@ucla.edu}
\altaffiltext{4}{Department of Physics and Astronomy, University of California, Riverside, CA 92521}
\altaffiltext{5}{Department of Physics, University of Wisconsin-Milwaukee, P.O. Box 413, Milwaukee, WI 53201}

\begin{abstract}

We analyze rest-frame optical morphologies and gas-phase kinematics as traced by rest-frame far-UV and optical spectra for a sample of 204
star forming galaxies in the redshift range $z\sim2-3$ drawn from the Keck Baryonic Structure Survey (KBSS).
We find that spectroscopic properties and gas-phase kinematics are closely linked to morphology:
compact galaxies with semi-major axis radii $r \lesssim 2$ kpc are substantially more likely than their larger counterparts to exhibit \lya\ in emission.  Although \lya\ emission
strength varies widely within galaxies of a given morphological type, all but one of 19 galaxies with \lya\ equivalent width
$W_{\lya} > 20$ \AA\ have compact and/or multiple-component morphologies with $r \leq 2.5$ kpc.
The velocity structure of absorption lines in the galactic continuum spectra also varies as a function of morphology.
Galaxies of all  morphological types drive similarly strong outflows (as traced by the blue wing of interstellar absorption line features), but the outflows of larger galaxies are less highly
ionized and exhibit larger optical depth at the systemic redshift that may correspond to a decreasing efficiency of feedback in evacuating gas from the galaxy.
This $v\sim0$ \kms\ gas is responsible both for shifting the mean absorption line redshift
and attenuating $W_{\lya}$ (via a longer resonant scattering path) in galaxies with larger rest-optical half light radii.
In contrast to galaxies at lower redshifts, there is no evidence for a correlation between outflow velocity and inclination, suggesting that outflows from
these puffy and irregular systems may be poorly collimated.
Our observations are broadly consistent with theoretical models of inside-out growth of galaxies in the young universe, in which 
typical $z\sim 2-3$ star forming galaxies are predominantly unstable, dispersion-dominated, 
systems fueled by rapid gas accretion that later form extended rotationally-supported disks when stabilized by a sufficiently massive stellar component.

\end{abstract}

\keywords{galaxies: high-redshift ---  galaxies: fundamental parameters --- galaxies: structure}

\section{INTRODUCTION}

Galaxy evolution is a process driven largely by the lifecycle of gas and its accretion, conversion into stars, and eventual loss via supernova-driven winds.
During the peak epoch of galaxy formation ($z\sim2-3$; \citealt{dickinson03,reddy08,zhu09})
those systems that dominate the luminosity function are especially gas-rich,
with gas frequently accounting for 50\% or more of the total baryonic mass (e.g., \citealt{erb06c,tacconi10}).
This large supply of hydrogen fuels sustained star formation rates (SFR) commonly 
$\sim 30 \, M_{\odot}$ yr$^{-1}$ (e.g., \citealt{erb06b,wuyts11})
and SFR surface densities peaking above 
$\sim 1 \, M_{\odot}$ yr$^{-1}$ kpc$^{-2}$ (e.g., \citealt{genzel11})
that are comparable to those in transient starbursts in the modern universe (\citealt{kennicutt98b}).

The mechanisms by which these galaxies acquire such large quantities of gas are uncertain;
although gas has traditionally been assumed to accrete  primarily through
mergers (e.g., \citealt{robertson06}) and `hot mode' accretion via virial shocks (e.g., \citealt{rees77,white78,white91}), some recent arguments suggest that 
semi-continuous cold accretion from cosmological
filaments  (e.g., \citealt{keres05,keres09,dekel09,be09,ceverino10})  may be required to sustain the observed
SFR and produce the structures observed at $z\sim2$.
As discussed by Reddy et al. (2012) however, star formation at $z\sim2-3$ appears to have been an inefficient process, indicating that SFR
is unlikely to have been limited primarily by the cold gas accretion rate.

In contrast to galaxies in the nearby universe, $z\sim2$ star forming galaxies do not exhibit regular
spiral structure (with some notable exceptions; \citealt{law12b}) and instead tend to be compact, triaxial systems consisting of one or more 
irregularly-shaped clumps of emission (e.g., \citealt{conselice05,elmegreen05,ravindranath06,law12a}; and references therein).
In many of the most massive galaxies ($M_{\ast} \gtrsim 3 \times 10^{10} M_{\odot}$) such clumps may be embedded in low surface-brightness structures reminiscent
of thick disks with typical scaleheights $h_z \sim 1$ kpc (e.g., Elmegreen \& Elmegreen 2006; Genzel et al. 2006, 2008;
 F{\"o}rster Schreiber et al. 2011a, 2011b) sustained by their large vertical velocity dispersions (e.g., Genzel et al. 2006; Law et al.  2007, 2009; F{\"o}rster Schreiber et al. 2009;
 Swinbank et al. 2011).

The large cold gas fractions of these galaxies appears to play a significant role in their dynamical
and morphological evolution.  Simulations (e.g., Noguchi 1999; 
Immeli et al. 2004; Bournaud et al. 2007; Bournaud \& Elmegreen 2009) indicate that such gas-rich galaxies should be highly unstable to fragmentation, forming clumps similar
to those observed which may either migrate to form a central bulge (Immeli et al. 2004; Elmegreen et al. 2008; Ceverino et al. 2010)
or be rapidly disrupted by stellar feedback (Wuyts et al. 2012).
A reliable understanding of the relation between galactic stellar morphology and its gas-phase properties is therefore critical for constraining models of galaxy formation
in the young universe.

While observational signatures of {\it infalling} gas (whether hot or cold-mode) have proven subtle and difficult to detect
(e.g., van de Voort \& Schaye 2012; although c.f. Rauch et al. (2011) and \citealt{kulas12}), the intense star formation of the galaxies gives rise to ubiquitous
and well-studied galactic-scale {\it outflows} evident in rest-frame UV spectra
(e.g., Shapley et al. 2003; Weiner et al. 2009; Steidel et al. 2010).  These outflows are frequently powerful enough to 
drive enriched gas to the virial radius and beyond with mass loss rates comparable to or exceeding the star formation rate (\citealt{pettini00,erb08,steidel10}).
Numerous authors (e.g., Shapley et al. 2003, Penterrici et al. 2010; Kornei et al. 2010; Steidel et al. 2010; Jones et al. 2012)
have therefore sought to use the information about optical depth, covering fraction, geometry, and gas kinematics encoded within such spectra
to learn about the mechanisms by which cold gas is coupled to feedback and physical observables such as
stellar mass, star formation rate, and stellar population age.
Efforts to understand the role of the gas in shaping the evolution of galaxy morphology however (e.g., Law et al. 2007b; Penterrici et al. 2010) have historically been 
limited in their ability to test current galaxy formation models by the lack of availability of high resolution rest-frame optical imaging tracing the bulk of the stellar mass.

In the present contribution, we combine rest-frame optical imaging of a large sample of $z\sim2$ star forming galaxies recently obtained with the WFC3 camera on board
the {\it Hubble Space Telescope} ({\it HST}) with ground-based rest-UV spectroscopy to investigate the relation between galaxy morphology and gas-phase kinematics.
In Section \ref{obs.sec} we describe the target galaxy sample and detail the methods by which morphological, photometric, and  spectroscopic  data were obtained.
We discuss the Ly$\alpha$ emission properties of the sample in \S \ref{lya.sec}, 
and the structure of major interstellar absorption lines arising in outflowing cold gas in \S \ref{metals.sec}, highlighting the changing velocity substructure of the absorption
lines as a function of galaxy morphology.  We discuss the implications of our findings for the evolution of typical $z\sim2-3$ star forming galaxies in \S \ref{discussion.sec},
combining our results with recent observations from IFU kinematic surveys.  We summarize our conclusions in \S \ref{summary.sec}.

Throughout our analysis, we adopt a standard $\Lambda$CDM cosmology based on the seven-year WMAP results (Komatsu et al. 2011) in which $H_0 = 70.4$ km s$^{-1}$ Mpc$^{-1}$, $\Omega_M = 0.272$,
and $\Omega_{\Lambda} = 0.728$.

\section{OBSERVATIONAL DATA}
\label{obs.sec}

\begin{figure*}
\plotone{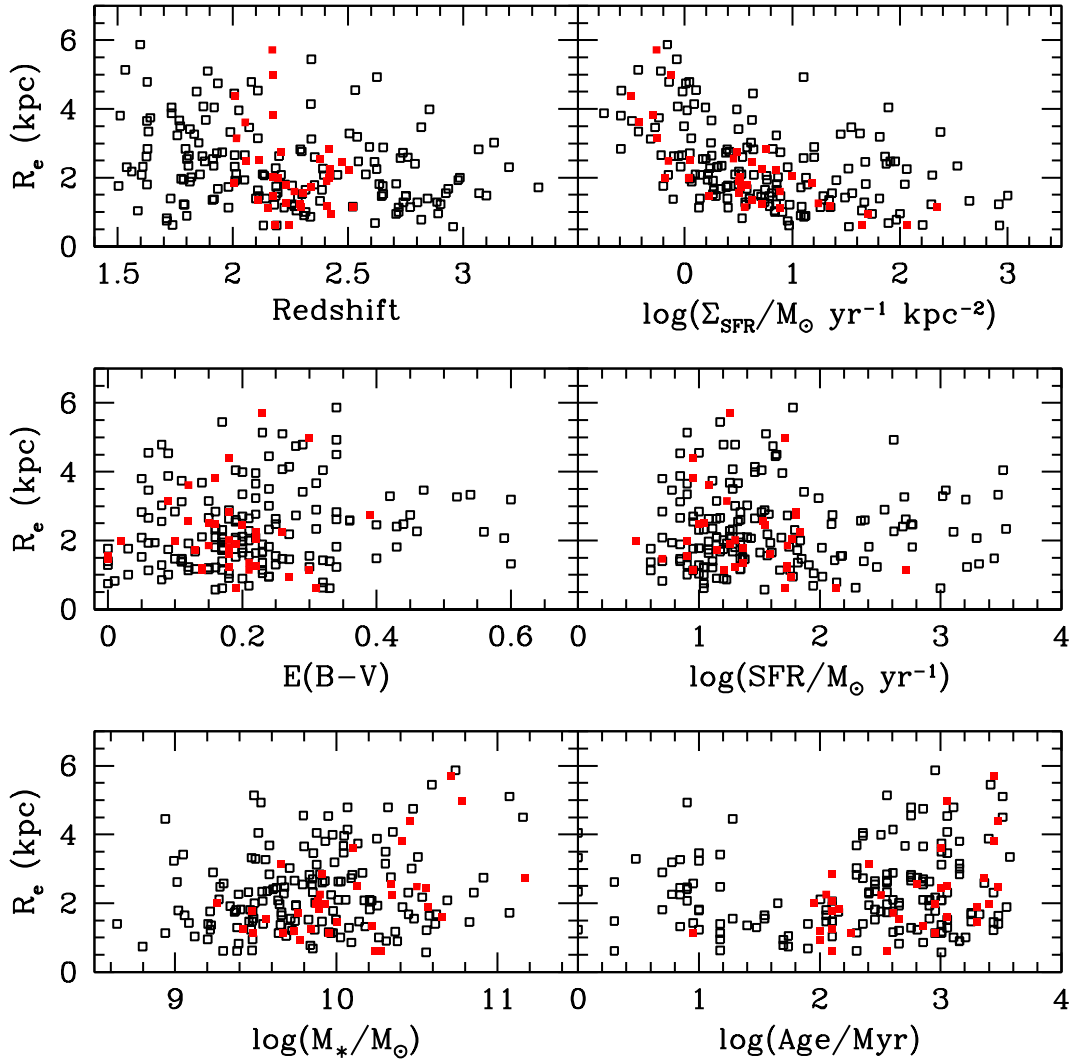}
\caption{Redshifts, stellar population parameters, and effective radii derived for the parent (black open boxes) and H$\alpha$ high-quality (HAHQ; red filled boxes) samples of 204 and 35 galaxies respectively.  The two samples have significant overlap, but the HAHQ sample is the product of a complicated selection function resulting in
a more restricted redshift range and marginally larger stellar masses and population ages
than the parent population on average.}
\label{parmrange.fig}
\end{figure*}

Our galaxy sample is drawn from the Keck Baryonic Structure Survey (KBSS; Trainor \& Steidel 2012), a catalog of 
$z \sim 1.5-3.6$ star forming galaxies selected according to optical $U_n G {\cal R}$ color (\citealt{steidel03,steidel04,adelberger04})
and confirmed using rest-UV spectroscopy.  These galaxies typically lie near sightlines to hyperluminous
background QSOs  ($z_{\rm QSO} \sim 2.7$) scattered widely across the sky.
We define two specific samples of galaxies used in our analysis:

\begin{description}
\item [Parent sample:]  A sample of 204 galaxies with {\it HST}/WFC3 imaging data, magnitudes $H_{160} \lesssim 24$ AB, and
redshifts $z \sim 1.5-3.5$ (see Fig. \ref{parmrange.fig}) derived from rest-UV spectroscopy.
\item [HAHQ subsample:] A subgroup of 35 galaxies from the parent sample for which the rest-UV spectroscopy is particularly high-quality
(`uvqual' $\geq 1$; see \S \ref{uvspectra.sec})
and secure systemic redshifts in the range $z=2.0 - 2.5$ (see Fig. \ref{parmrange.fig}) have been obtained from H$\alpha$ nebular emission line spectroscopy.
\end{description}

\subsection{Rest-Optical Morphologies}

As part of {\it Hubble Space Telescope} ({\it HST}) Cycle 17 program GO-11694, we obtained WFC3/IR imaging of
306 $U_n G {\cal R}$ color-selected star forming galaxies with spectroscopic redshifts in the range $z=1.5-3.6$
drawn from the KBSS.
The details of these observations have been described at length by Law et al. (2012a; hereafter `Paper I').
In brief, we used the F160W ($\lambda_{\rm eff} = 15369$ \AA) filter to trace rest-frame optical emission ($\lambda \sim 3400-6100$ \AA\ for the parent sample,
$\lambda \sim 4400-5100$ \AA\ for the HAHQ sample).
Our data reach a depth of 27.9 AB for a $5\sigma$ detection within a 0.2 arcsec radius aperture (representing a 
total of 8100s integration in each pointing), and have a PSF FWHM of 0.18'' that corresponds to 1.5 kpc at $z \sim 2$.

The morphologies consist of three qualitative types.  As defined in Paper I, these are
\begin{description}
\item [Type I:] Single nucleated component with little to no
low surface brightness emission.
\item [Type 2:] Two or more distinct nucleated components of comparable magnitude and little
to no low surface brightness emission.
\item [Type 3:] Extended objects with 
non-axisymmetric low surface-brightness features.
\end{description}

Numerous morphological statistics have been defined in the literature that attempt
to encapsulate this range of morphologies quantitatively.
In Paper I we calculated 7 of the most widely used such statistics for each of the galaxies in our WFC3 imaging fields.  These statistics are:

\begin{description}
\item[$r$:] Semi-major axis radius derived from fitting Sersic (1963) models convolved with the observational PSF to the F160W light profile using the GALFIT (Peng et al. 2002, 2010) analysis package.  We also define the circularized effective radius ($r_{\rm e} = r \sqrt{b/a}$) for consistency with previous analyses, where $b/a$ is the apparent
minor/major axis ratio.
\item[$n$:] Radial index of the best-fit Sersic model profile.
\item[$G$:] Gini statistic (Abraham et al. 2003; Lotz et al. 2004) quantifying the cumulative flux distribution among the galaxy pixels.
\item[$M_{20}$:] Second order moment of the spatial distribution of the light profile (Lotz et al. 2004, 2006).
\item[$C$:] Concentration index of the flux distribution about the center of the galaxy (Bershady et al. 2000; Conselice 2003).
\item[$A$:] Asymmetry index (Schade et al. 1995; Conselice et al. 2000) quantifying the $180^{\circ}$ rotational asymmetry of the galaxy.
\item[$\Psi$:] Multiplicity index (Law et al. 2007b) quantifying the `work' required to assemble the distribution of pixel fluxes within a galaxy (distinguishing single-component from 
irregular and multiple-component sources).
\end{description}

Although the first two morphological statistics ($r,n$) assume a parametric model convolved with the observational PSF, the latter five ($G, M_{20}, C, A, \Psi$)
are non-parametrically derived from the spatial arrangement of individual pixel fluxes.  The values calculated for these five statistics depend systematically
on the adopted segmentation map defining which pixels are considered to be part of a given galaxy.  Our approach is placed on the systematic reference frame of
Lotz et al. (2008a, 2008b, and references therein) using the transformations described in Appendix A.1 of Paper I.

All seven morphological statistics are robust for galaxies with $H_{160} \lesssim 24$ AB, and show systematic bias
at fainter magnitudes (see Appendix A.2 of Paper I).
We therefore restrict our attention in the present contribution to those 204 galaxies with $H_{160} \leq 24.0$ and no evidence of QSO or AGN in their 
UV spectra (\S \ref{uvspectra.sec}) or broadband photometry (\S \ref{sed.sec}).
As derived in Paper I, typical uncertainties for individual galaxies are $\sim$ 2\% in $r$, 15\% in $n$, 3\% in $G$, 4\% in $M_{20}$, 11\% in $C$, 22\% in $A$, and 21\% in $\Psi$.

\subsection{Rest-UV Spectroscopy}
\label{uvspectra.sec}

Rest-frame UV spectra for each of our galaxies have been obtained using the LRIS multislit spectrograph on Keck I.  
Details concerning the observational setup, data reduction, and calibration process of these spectra
have been presented at length elsewhere in the literature (e.g., Shapley et al. 2003; Steidel et al. 2010; and references therein).  
In general, individual spectra represent integration times $\sim$ 1.5 hours 
with characteristic velocity resolution $\sigma_{\rm res} \approx 160$ \kms .
These spectra vary considerably in quality
according to both the intrinsic brightness of the target galaxy and the conditions under which particular slit masks were observed.
We sort the galaxies according to the quality of their spectra, assigning them quality indices `uvqual' ranging from 0 (one or two spectroscopic features identified with
relatively low confidence) to 3 (multiple strong spectroscopic features identified with extremely high confidence).


The limited S/N ratio of even the highest-quality spectra however means that it is typically only possible to estimate the 
mean interstellar absorption (emission) line redshift $z_{\rm IS}$ ($z_{\rm EM}$) for individual 
galaxies.  In order to investigate the detailed structure of spectral lines it is necessary to stack together multiple spectra to reach higher S/N ratios.
This stacking process is described in detail by Steidel et al. (2010), and includes doppler-correcting each spectrum to the rest frame,\footnote{Where available we use
the nebular emission line redshift derived from rest-optical spectroscopy as detailed in \S \ref{optspec.sec}, otherwise we adopt the prescriptions given by Steidel et al. (2010).}
resampling to a common 0.33 \AA/pixel spectral dispersion, and
rescaling to a common mode in the wavelength range $\lambda \sim 1250-1500$ \AA\ before combining individual spectra using a sigma-clipped mean algorithm.

\begin{figure*}
\plotone{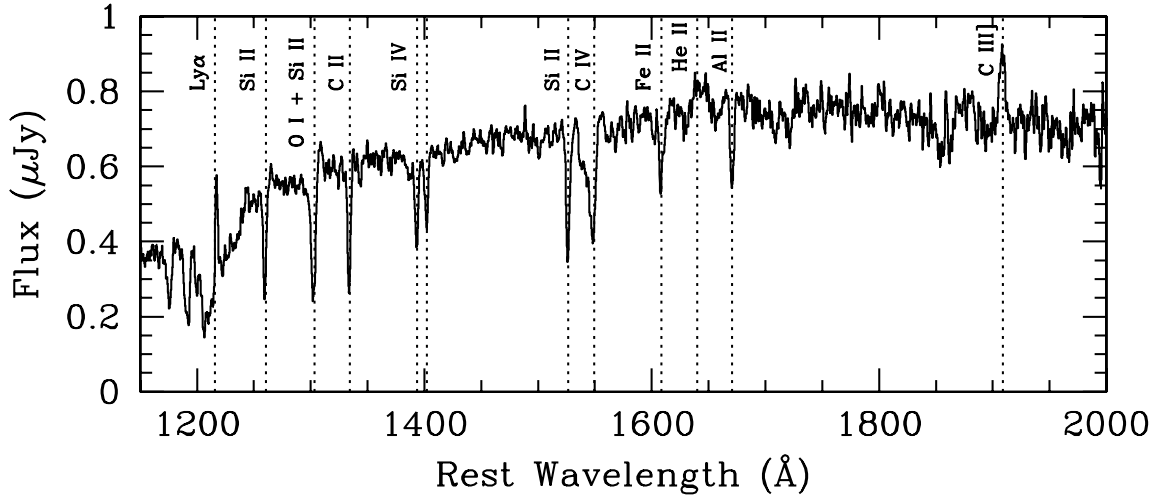}
\caption{Flux-calibrated rest-UV Keck/LRIS-B spectrum constructed from stacking 204 star-forming galaxies in our parent sample in the redshift range $1.5 < z < 3.5$.  
Vertical dotted lines indicate the fiducial wavelengths of major emission and/or absorption line transitions.}
\label{specstack.fig}
\end{figure*}

In Figure \ref{specstack.fig} we show the composite spectrum for the 204 galaxies in our parent sample.
This composite spectrum exhibits a mixture of Ly$\alpha$ absorption and emission, deep blended interstellar absorption features due to various metal species in the galactic ISM,
and a host of weaker  features due to stellar photospheric absorption and fine structure emission.
These absorption features are typically  blueshifted with respect to the systemic redshift (corresponding to absorption arising in the near side of an outflowing galactic wind),
while Ly$\alpha$ emission is redshifted (corresponding to off-resonance backscattering from the far side of the outflow).
As described at length by Steidel et al. (2010; see also Kulas et al. 2012), this outflow is likely composed of discrete gas clouds whose expansion
velocities increase with galactocentric radius.

\subsection{Rest-optical Spectroscopy}
\label{optspec.sec}

Although the rest-UV spectra give an indication of the redshift of the interstellar absorption lines $z_{\rm IS}$, it is not possible to perform a detailed analysis
of the kinematic structure of the outflowing gas without knowing the systemic redshift $z_{\rm sys}$ of the galaxy as well.
We therefore define the  `H$\alpha$ - high quality' (HAHQ) subsample of galaxies with particularly robust measurements of $z_{\rm IS}$ and for which
accurate (uncertainty $\sim 60$ \kms; Steidel et al. 2010) systemic redshifts have been derived from H$\alpha$ nebular emission
using a combination of long-slit Keck/NIRSPEC (Erb et al. 2006b), integral-field Keck/OSIRIS (Law et al. 2007a, 2009; Wright et al. 2009) and 
VLT/SINFONI (F{\"o}rster Schreiber et al. 2009)  spectroscopy.
The HAHQ sample represents the subset of 35 of the 89 galaxies considered by Steidel et al. (2010) for which {\it HST}/WFC3 morphologies have been obtained.

\subsection{Stellar Population Models}
\label{sed.sec}

Extensive KBSS ancillary data exist for our target galaxies, typically including 
deep ground-based optical and infrared $J/K_s$ photometry, and in some cases {\it Spitzer} IRAC and/or MIPS photometry.  
In combination with $H_{160}$ magnitudes derived from the {\it HST}/WFC3 imaging data, the long-wavelength photometry permits
us to construct stellar population models by fitting the broadband spectral energy distribution of each galaxy as detailed
in Paper I (see also Shapley et al. 2001, 2005; Erb et al. 2006c;
Reddy et al. 2006, 2010).  In brief, we fit Charlot \& Bruzual (2012, in prep.) stellar population synthesis models in combination with a Chabrier (2003) IMF,
Calzetti et al. (2000) extinction law,
and a constant ($\tau = \infty$) star formation history using a customized IDL code (Reddy et al. 2012).
Monte Carlo tests for the 204-galaxy parent sample indicate that the mean fractional uncertainties in derived values for $E(B-V)$, age, SFR, and $M_{\ast}$ are
$\langle \sigma_x / x \rangle = $ 0.3, 0.6, 0.5, and 0.4 respectively (consistent with previous measurements for the larger KBSS sample by Shapley et al. 2005, Erb et al. 2006b,
and Reddy et al. 2012).

As illustrated in Figure \ref{parmrange.fig}, galaxies in the parent sample have stellar masses in the range $M_{\ast} \sim 10^9 - 10^{11} \, M_{\odot}$, star formation rates
$\sim 30 \, M_{\odot}$ yr$^{-1}$, and moderate color excess $E(B-V) \sim 0.0 - 0.4$.  The HAHQ subsample is biased relative to the parent sample
by the requirement of both detecting H$\alpha$ emission and observing a high-quality UV spectrum (favoring UV-bright objects), 
and therefore has a slightly larger mean stellar mass ($\langle {\rm log}\, M_{\ast}/M_{\odot} \rangle_{\rm HAHQ} = 10.1 \pm 0.4$
vs. $\langle {\rm log}\, M_{\ast}/M_{\odot} \rangle_{\rm Parent} = 9.9 \pm 0.5$)\footnote{$1\sigma$ values represent standard deviation about the mean.}.

\section{Ly$\alpha$ Emission}
\label{lya.sec}

Ly$\alpha$ emission originates deep within individual star forming regions and  is resonantly scattered throughout the interstellar medium 
until either being absorbed on a dust grain or redshifted sufficiently far off-resonance that it can escape the galaxy.  Although almost all star forming galaxies appear to
have Ly$\alpha$ emitting haloes on large angular scales (\citealt{steidel11}), the  Ly$\alpha$ profile of the central few kpc traced by slit spectra
is dependent upon the structure, kinematics, and ionization properties of each galaxy and its interstellar and circumgalactic medium.
Substantial efforts have been made over the last decade to attempt to disentangle these various effects and connect the central Ly$\alpha$ line profile 
with global stellar population properties
both for narrowband-selected Ly$\alpha$-Emitters (LAEs; e.g., Gawiser et al. 2007) and for star forming galaxies selected via broadband optical color.
\cite{shapley03}, for instance, found that $z\sim3$ Lyman Break Galaxies (LBGs) with
stronger \lya\ emission tend to have 
lower SFR, bluer UV continua, and weaker low-ionization interstellar absorption lines.
Similar results were found for $z\sim2$ star forming galaxies by \cite{erb06a} and \cite{law07b}, who noted that 
galaxies with stellar masses $M_{\ast} > 10^{10} M_{\odot}$ on average have weaker \lya\ emission than do galaxies with $M_{\ast} < 10^{10} M_{\odot}$,
and at $z\sim4$ by Jones et al. (2012).
These results were borne out by the more detailed analysis of \cite{steidel10}, which demonstrated that \lya\ 
emission is both weaker and more strongly redshifted from the systemic velocity
in galaxies with high baryonic mass, consistent with the picture that
\lya\ emission from the deepest gravitational potential wells must scatter further from resonance
before being able to escape.
Using a similarly selected sample of LBGs at $z=2.5 - 3.5$, \cite{penterrici10} found comparable results, showing that galaxies without \lya\ emission
tend to be more massive and dustier, although these authors observed that stellar population ages 
and SFRs  did not depend strongly on \lya\ emission characteristics.

The relation between \lya\ emission and rest-frame UV galaxy morphology has previously been investigated by \cite{law07b} and \cite{penterrici10} (see also Bond et al. 2009, Malhotra et al. 2012).
\cite{law07b}  found that \lya\ emission was correlated with the Gini coefficient in the sense that high-$G$ (i.e., strongly nucleated) systems were more likely to exhibit \lya\
emission, which we previously ascribed to the lower quantity of dust in strongly nucleated galaxies.
Although \cite{penterrici10} found no significant correlation of \lya\ emission with
any of the non-parametric morphological indices  ($C, G, A$, or $M_{20}$), these authors noted that while galaxies with weak or no \lya\ emission spanned a range
of projected angular sizes, galaxies with strong \lya\ emission tended to be exclusively small.

In \S \ref{lya_prob.sec} we discuss the probability that galaxies within our sample exhibit \lya\ emission as a function of various physical properties, and expand our discussion
in \S \ref{lya_strength.sec} to consider the trends in emission line strength within the \lya-emitting subset.

\subsection{Probability of Ly$\alpha$ Emission}
\label{lya_prob.sec}

We divide the parent sample into two 
groups according to whether  \lya\ emission is present (71 galaxies)
or not (117 galaxies) in the UV spectrum of each galaxy.\footnote{We rejected 16 galaxies in the redshift range 
$z=1.5-1.7$ from our sample since \lya\ emission at these redshifts does not fall within the LRIS UV spectral range.}
The first of these two groups includes 
galaxies with \lya\ emission of any strength, including those with a small emission peak superimposed on a large \lya\ absorption trough.
We later discuss the properties of a `\lya-bright' subgroup which meets the $W_{\lya} > 20$ \AA\ selection criterion used by dedicated surveys
of \lya-emitters (LAEs; e.g., Gawiser et al. 2007).

In Table \ref{lyacorrel.tab} we list probabilities calculated using 
the nonparametric Kolmogorov-Smirnov (KS) test to evaluate whether the two groups are statistically consistent with the null hypotheses
that they are drawn from the same distribution of stellar mass, SFR, morphology, etc.
We find that SFR and the morphological parameters $M_{20}$, $C$, $A$, $\Psi$, and $n$ are poor predictors of the likelihood of observing \lya\ emission from a galaxy;
galaxies with and without \lya\ in emission
have statistically indistinguishable distributions.
In contrast, the null hypothesis is rejected at greater than 94-95\% confidence
when we examine the mass, age, color excess, $G$, or $r$ distributions for galaxies both with and without \lya\ in emission.
These correlations 
are in the sense that galaxies with \lya\ emission on average are smaller, may have slightly lower mass  
($\langle {\rm log} (M_{\ast}/M_{\odot}) \rangle = 9.82$ vs $9.96$),
and slightly younger ages (500 vs 800 Myr), and are less dusty ($\langle E(B-V) \rangle = 0.18$ vs 0.23) than galaxies with no \lya\ emission.
These results are broadly consistent with relations that have already been established
by \cite{shapley03}, \cite{erb06a}, \cite{law07b}, \cite{steidel10}, and \cite{penterrici10}.

\begin{deluxetable}{lc}
\tablecolumns{2}
\tablewidth{200pt}
\tabletypesize{\scriptsize}
\tablecaption{KS probability that galaxies with and without Ly$\alpha$ emission are drawn from different distributions\tablenotemark{a}}
\tablehead{
\colhead{Quantity} & \colhead{$P_{\rm KS}$}}
\startdata
$M_{\ast}$ \tablenotemark{b} & {\bf 6\%} \\
Age \tablenotemark{c} & {\bf 8\%} \\
SFR \tablenotemark{d} & 99 \% \\
SigmaSFR \tablenotemark{e} & {\bf 1 \%} \\
E(B-V) \tablenotemark{f} & {\bf 3\%} \\
$G$ & {\bf 8\%} \\
$M_{20}$ & 83\% \\
$C$ & 26\%\\
$A$ & 99\% \\
$\Psi$ & 56 \% \\
$r$ \tablenotemark{g} & {\bf 3\%} \\
$r_{\rm e}$ \tablenotemark{h} & {\bf 0.5\%} \\
$n$ & 15 \% \\
$b/a$ \tablenotemark{i} & 59 \% \\
$H_{160}$ & 12 \% \\
\enddata
\tablenotetext{a}{All values are one minus the probability at which the null hypothesis (that the quantities are drawn from a
statistically identical distribution of the indicated parameter) is ruled out.  The sample size is 188 galaxies.}
\tablenotetext{b}{Stellar mass, estimated from SED fitting.}
\tablenotetext{c}{Average population age from a constant SFR SED model.}
\tablenotetext{d}{Star formation rate measured from SED fitting.}
\tablenotetext{e}{Average star formation rate surface density $\Sigma_{\rm SFR} = SFR/ \pi r_{\rm e}^2$.}
\tablenotetext{f}{Color excess due to dust extinction from a constant SFR SED model.}
\tablenotetext{g}{GALFIT semi-major axis radius.}
\tablenotetext{h}{GALFIT circularized effective radius.}
\tablenotetext{i}{Morphological minor/major axis ratio.}
\label{lyacorrel.tab}
\end{deluxetable}

We note, however, that \cite{kornei10} studied the Ly$\alpha$ properties of $z\sim3$ LBGs and found a conflicting result that 
those with $W_{\lya} > 20$ \AA\  tended to have stellar
populations slightly {\it older} than their non-emitting kin, which was hypothesized to indicate that \lya\ emitters may represent a later stage in galaxy evolution in which
supernova-driven outflows have reduced the dust covering fraction (see also Shapley et al. 2001).\footnote{\cite{kornei10} used an equivalent width measure
$W_{\lya} > 20$ \AA\ to define their \lya-emitting population, in contrast to our simple criterion that there be any measurable \lya\ emission.  However, our results
are qualitatively unchanged if we instead consider the \lya-bright subset of galaxies with  $W_{\lya} > 20$ \AA.}
The difference with our present findings may partly reflect the different mean redshift range of the samples, or may indicate that 
the relation between \lya\ emission strength and stellar population age is simply too weak (96\% rejection of the null hypothesis)
to measure consistently given typical uncertainties in the derived ages of individual galaxies.  Indeed, while \cite{penterrici10} concurred 
that \lya\ emitters tended to be slightly younger than similar-selected non-emitters on average, these authors too noted that the correlation was only marginally significant.

As indicated by Table \ref{lyacorrel.tab}, within our sample of star forming galaxies 
there is no statistically significant correlation between the presence/absence of \lya\ emission and the morphological indices
$C$, $A$, $M_{20}$, $\Psi$, or $n$.  Similar to the result obtained by \cite{law07b}, 
there is a relation (92\% confidence) between \lya\ emission and the Gini parameter $G$
in the sense that galaxies with \lya\ emission tend to have marginally more strongly nucleated light profiles in the rest-frame optical.
This trend does not appear to be of statistical significance however as 
the mean values of $G$  for galaxies with and without
\lya\ emission respectively differ by only $\sim 1\sigma$ ( $\langle G \rangle_{\lya} = 0.49 \pm 0.01$ vs $\langle G \rangle_{\rm No \lya} = 0.478 \pm 0.005$).\footnote{Here
the uncertainties represent the $1/\sqrt{N}$ uncertainty in the mean of each sample.}

\begin{figure*}
\plottwo{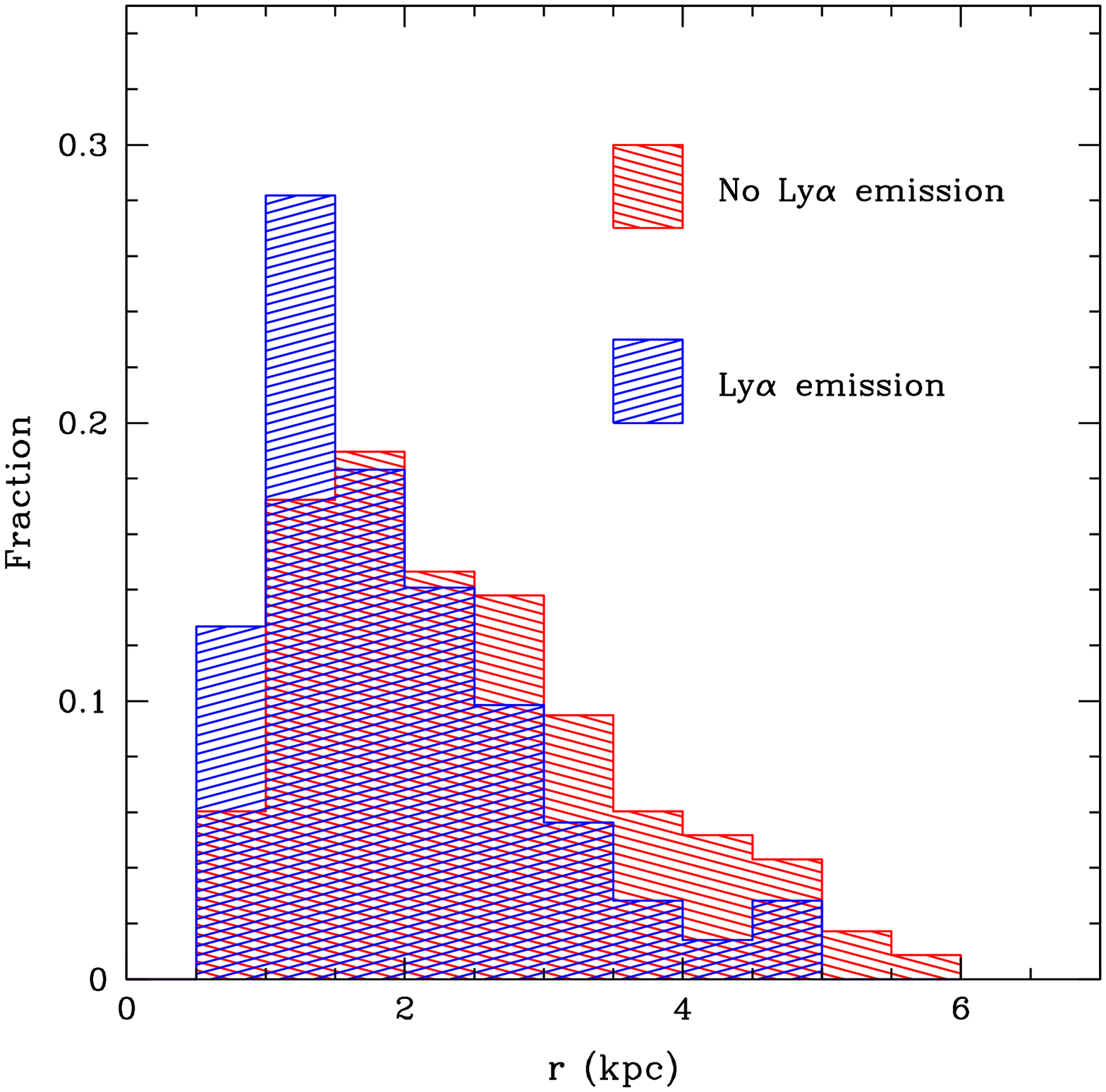}{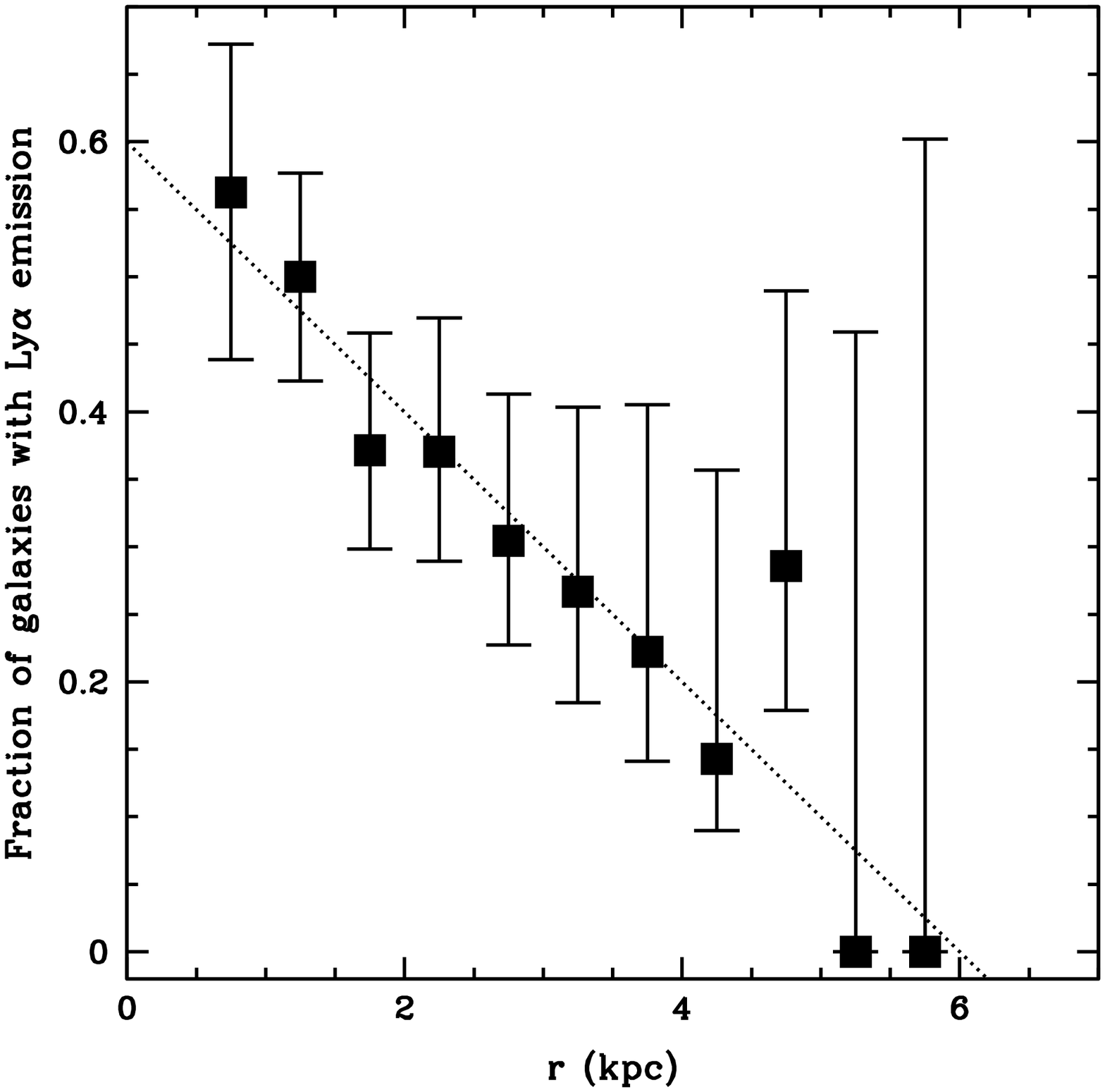}
\caption{Left panel: Histogram of semi-major axis radius $r$ for galaxies with and without  \lya\ emission component in the 183 galaxies of the parent sample for which the LRIS
spectral range includes the \lya\ transition.  Fractions shown are relative to the total number of galaxies with (71 galaxies) and without (117 galaxies) \lya\ emission respectively.  Right panel:
Fraction of galaxies in each radial bin that show \lya\ emission; $1\sigma$ uncertainties are estimated using a binomial probability distribution.  The dashed line indicates a least-squares fit of the form $y = 0.60 - 0.10 \times x$ to the data; a zero slope is disfavored with $\sim 3\sigma$ confidence.}
\label{rlyahist.fig}
\end{figure*}

In contrast, the presence of \lya\ emission is strongly correlated with the morphological semi-major axis radius $r$, circularized radius $r_{\rm e}$,
and (by construction) the star formation rate surface density $\Sigma_{\rm SFR} = {\rm SFR}/\pi r_{\rm_e}^2$.
These correlations are  significant at the 97-99\% confidence level, comparable
to or stronger than the well-known anti-correlation between \lya\ emission and dust content $E(B-V)$).
Similarly to the rest-UV angular size relation noted by \cite{penterrici10}, galaxies with \lya\ in emission
are more likely than non-emitters to have $r < 2.5$ kpc (74\% vs 56\%), and
the galaxies with $r < 2.5$ kpc in our sample are twice as likely (41\% vs 24\%) as those with $r > 2.5$ kpc to have Ly$\alpha$ in emission (see Figure \ref{rlyahist.fig}).  
This is unlikely to be due to any intrinsic correlation between semi-major axis radius and dust content; although $r$ and $E(B-V)$ of the parent sample are marginally
correlated at the $\sim 2\sigma$ level (based on the Spearman rank correlation test), this correlation is in the sense that larger galaxies tend to be marginally {\it less} dusty.

\begin{figure*}
\plotone{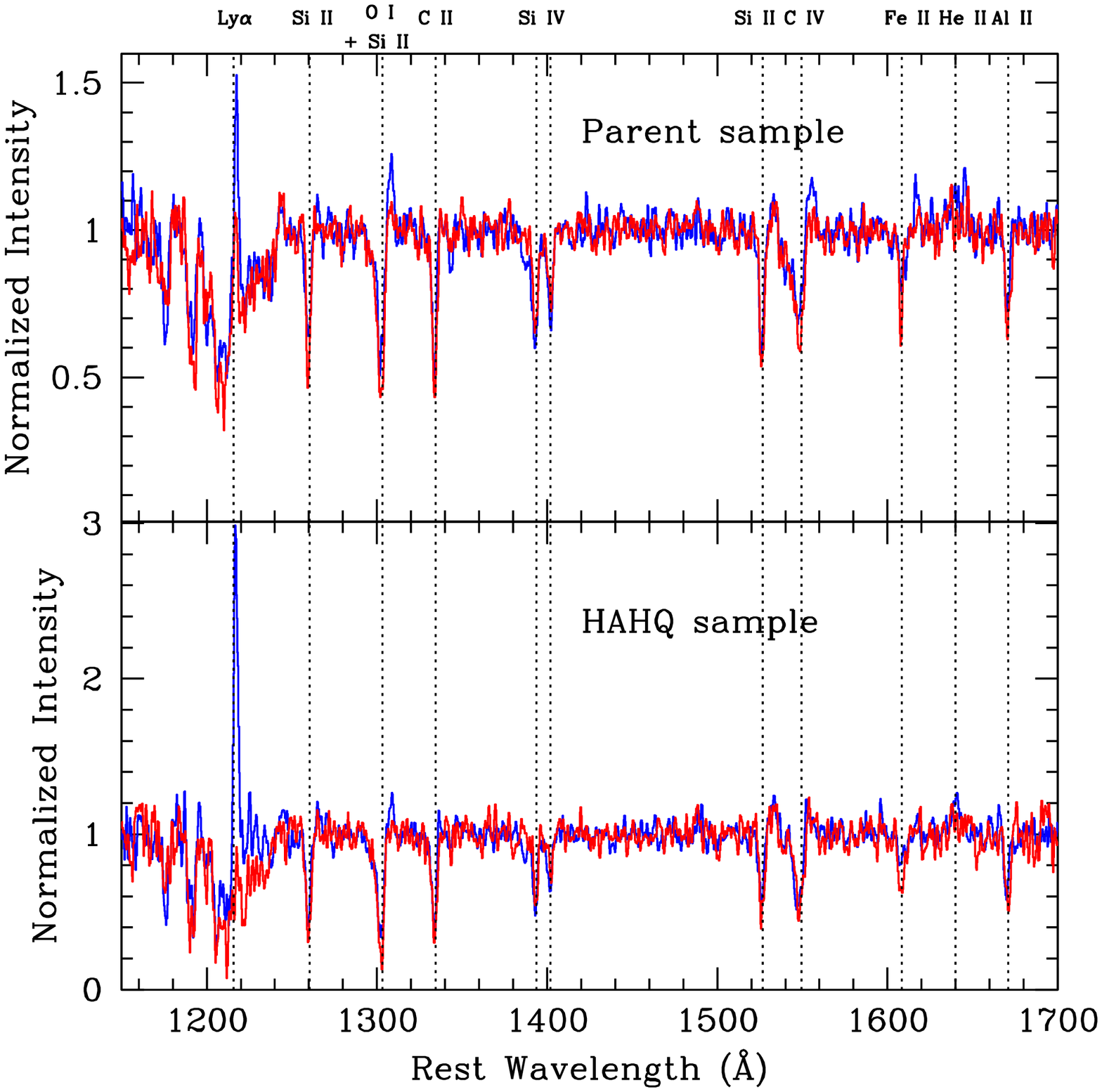}
\caption{Continuum-normalized stacked Keck/LRIS-B
UV spectra of the 50\% of galaxies with smallest (blue lines) and largest (red lines) sizes for the parent (200 galaxies) and HAHQ (35 galaxies) 
galaxy samples (note differing vertical scales).  The vertical dotted lines 
indicate the wavelengths of major spectral features arising from a combination of low/high ionization interstellar absorption and various emission mechanisms.  While not labeled, the
emission feature at $\sim$ 1300 \AA\ is the fine-structure emission line Si II${\ast} \lambda1309$\AA.  Note the drastic difference in Ly$\alpha$ profiles and varying depths of the low-ionization absorption features.}
\label{fullspec.fig}
\end{figure*}

We illustrate the spectroscopic difference between large- and small-$r$ galaxies by stacking the
halves of the parent and HAHQ samples with the smallest and largest semi-major axis radii 
(Figure \ref{fullspec.fig}).\footnote{The median radius used to divide the samples is 2.08 kpc and 1.98 kpc
for the parent and HAHQ samples respectively.}
The \lya\ emission equivalent width $W_{\lya}$ of the large and small-$r$ stacks is 1.3 \AA\ and 2.6 \AA\ respectively in the parent sample, representing a factor $\sim 2$ increase
in the mean \lya\ emission strength.\footnote{By comparison, Monte Carlo tests stacking randomly selected halves of the galaxy sample indicate that  $W_{\lya}$
of the stacks typically
vary by $\sim 30$\% for the parent sample, and by a factor $\sim 2$ for the HAHQ subsample.}
This difference is even more pronounced in the HAHQ subsample, increasing by a factor $\sim 10$ from the large-$r$ ($W_{\lya}$ = 0.7\AA)
to small-$r$ ($W_{\lya}$ = 6.7 \AA) stacks, corresponding to a K-S probability of $\sim 0.2$\% that the radii of galaxies with and without \lya\ in emission are drawn from the same distribution.

\subsection{Strength of Ly$\alpha$ Emission}
\label{lya_strength.sec}

Within the group of galaxies with \lya\ in emission  similar correlations exist between \lya\ emission {\it strength} and physical galaxy properties.
We compute the emission line equivalent width $W_{\lya}$ relative to the continuum level in our 
 flux-calibrated spectra for each of the 71 galaxies with such emission in the parent sample.\footnote{Based on an analysis of $\sim 1000$ galaxies with \lya\ emission
 in the KBSS, we estimate that uncertainties on $W_{\lya}$ for individual galaxies are typically $\sim 4$ \AA.}
In Table \ref{lyastrength.tab} we tabulate the correlations between $W_{\lya}$
and physical galaxy properties in terms of the number of standard deviations from the null
hypothesis (that the two variables are uncorrelated) using the Spearman rank correlation test.  The  factors most strongly influencing $W_{\lya}$ 
are the dust content $E(B-V)$ of a galaxy ($2.4\sigma$), its total star formation rate ($2.4\sigma$),\footnote{Although the SFR has no statistically significant
relation to the probability of observing \lya\ in emission, for the subsample of galaxies with \lya\ in emission we find the SFR to be correlated with the strength
$W_{\lya}$ of the emission.}
and its physical size ($2.3\sigma$), in the sense that
larger, dustier galaxies with higher SFR have lower $W_{\lya}$.

\begin{figure*}
\plotone{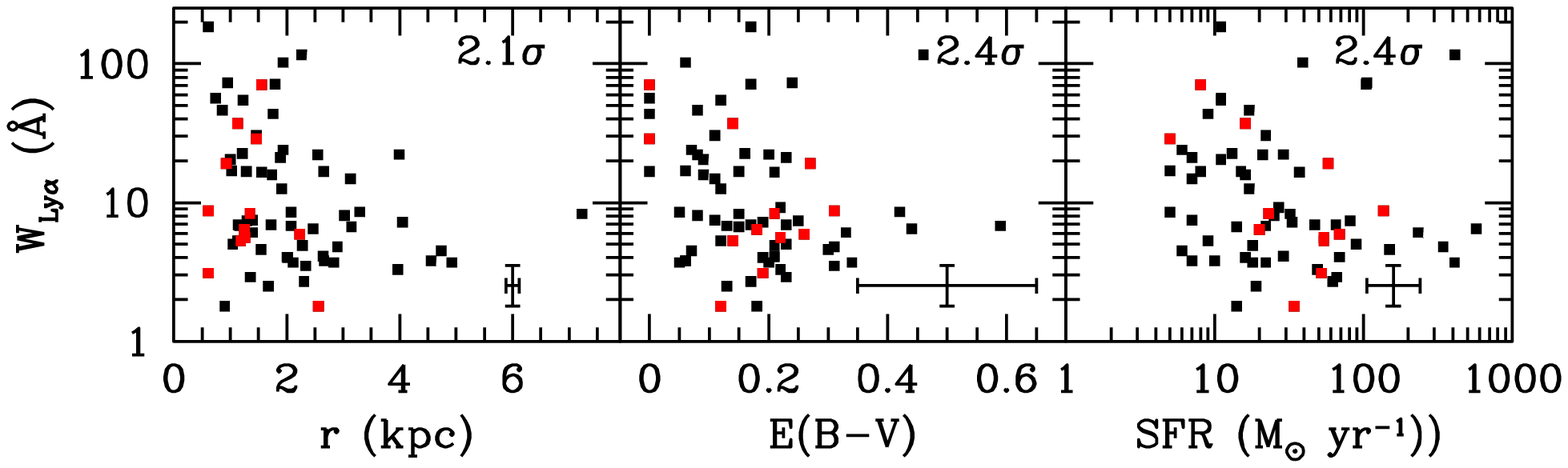}
\caption{Ly$\alpha$ emission equivalent width $W_{\lya}$ as a function of rest-optical semi-major axis radius $r$, dust content $E(B-V)$, and SED-derived star formation rate
 for the 71 galaxies with measurable \lya\ emission
in the parent sample (black points).  Red points denote the 12 \lya\ emitting galaxies that are also in the HAHQ subsample.   $W_{\lya}$ decreases with $r$,
$E(B-V)$, and SFR; the significance of the correlations as estimated from the Spearman rank correlation test are indicated in each panel.  Error bars in the lower-right corner of
each panel indicate the characteristic uncertainty of individual points.}
\label{lyastrength.fig}
\end{figure*}

\begin{figure*}
\plotone{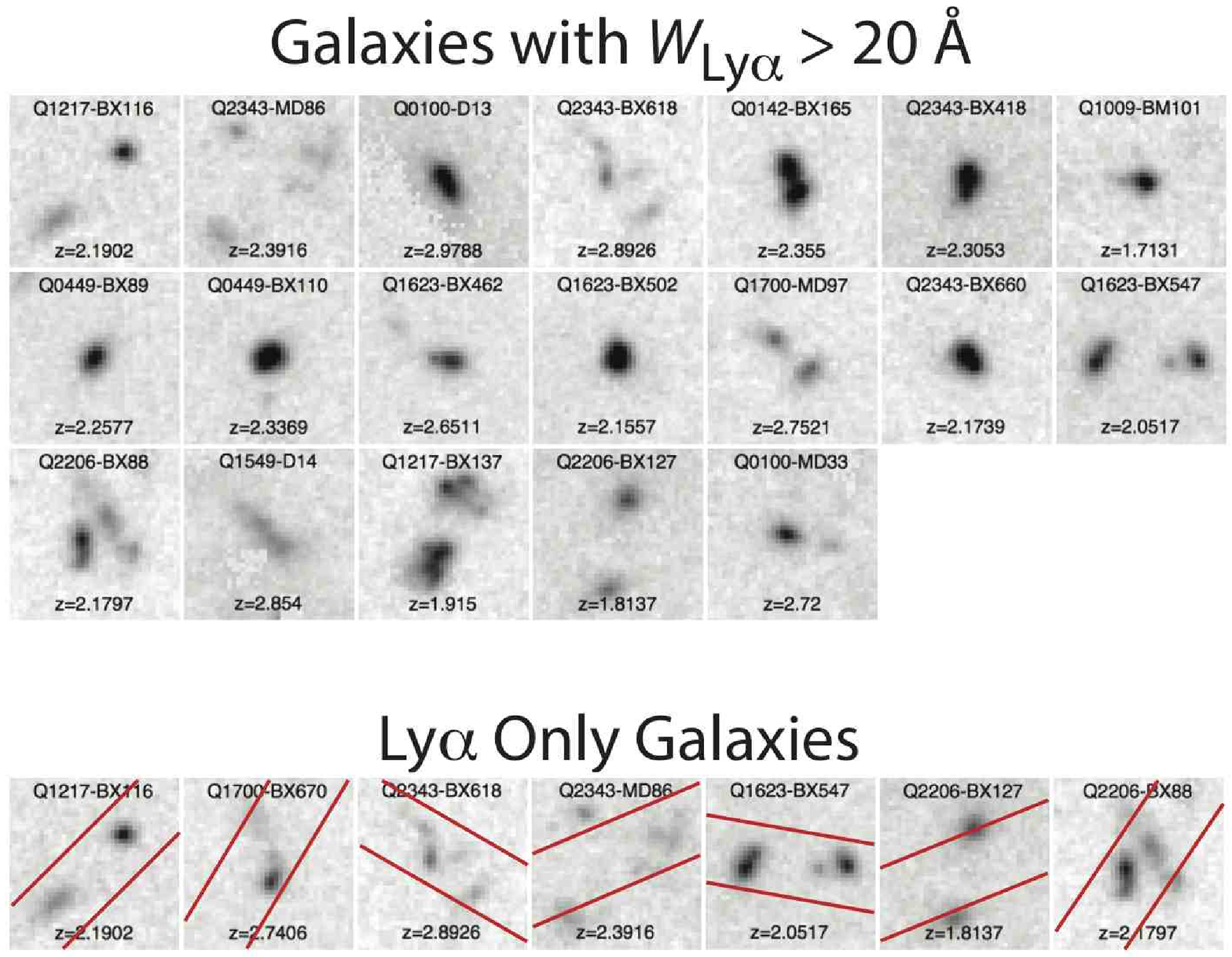}
\caption{{\it HST}/WFC3 F160W rest-optical morphologies of the 19 galaxies in the parent sample with \lya\ emission equivalent width $W_{\lya} > 20$ \AA\ (top panel), and
of the 7 galaxies in the parent sample with \lya\ emission but no absorption features in their rest-UV spectra.  Both sets of galaxies are
sorted in order of decreasing $W_{\lya}$).  Note that 18 of the 19 galaxies in the top panel are compact with Type I or II morphology, while all galaxies in the lower panel
have type II morphology.
Images are 3'' to a side, and oriented with North up and East to the left and centered on the F160W flux centroid.  The colormap has been inverted and uses an arcsinh stretch with
the black point set to 27.3 AB pixel$^{-1}$ (21.8 AB arcsec$^{-2}$).    Red lines in the lower panel
indicate the width and position angle of the LRIS-B slit during observations of each galaxy.}
\label{lyamorphs.fig}
\end{figure*}

We illustrate these three trends in Figure \ref{lyastrength.fig}.  
Similarly to previous studies (e.g., Shapley et al. 2003; Kornei et al. 2010; Penterrici et al. 2010), we find that galaxies with the strongest 
$W_{\lya}$ tend to contain lower than average quantities of dust and have lower star formation rates.
Additionally,  galaxies with large rest-optical
sizes ($r \gtrsim 3$ kpc) tend to have $W_{\lya} < 20$ \AA, while the smallest galaxies ($r \lesssim 2$ kpc) exhibit a much wider range in equivalent width from
$W_{\lya} = 0 - 180$ \AA.  Of the 19 star-forming galaxies in our optical-color selected sample with $W_{\lya} \gtrsim 20$ \AA\, all but one have rest-optical radii
$r \leq 2.5$ kpc and Type I and II morphologies (i.e., one or two nucleated components to the light profile with little to no extended and/or diffuse low surface brightness emission 
in the WFC3 imaging data; see Figure \ref{lyamorphs.fig}).
The correlation between \lya\ equivalent width and optical semi-major axis radius is independent of the correlation with $E(B-V)$; $r$ and $E(B-V)$ are consistent with
the null hypothesis of being uncorrelated ($0.1\sigma$ rejection of the null hypothesis using the Spearman rank correlation test), and confining our sample to a narrow range of values
in $E(B-V)$ results in a similarly strong trend of $W_{\lya}$ with $r$ after accounting for the reduced sample size.

\begin{deluxetable}{lc}
\tablecolumns{2}
\tablewidth{200pt}
\tabletypesize{\scriptsize}
\tablecaption{Correlation of $W_{\lya}$ and Galaxy Properties\tablenotemark{a}}
\tablehead{
\colhead{Quantity} & \colhead{Std. Dev.}}
\startdata
$M_{\ast}$\tablenotemark{b} & -1.7 \\
Age \tablenotemark{c} & 0.5 \\
SFR \tablenotemark{d} & -2.4 \\
$\Sigma_{\rm SFR}$ \tablenotemark{e} & -0.2\\
E(B-V) \tablenotemark{f} & {\bf -2.4} \\
$G$ & -0.2 \\
$M_{20}$ & 2.2 \\
$C$ & -0.7\\
$A$ & 1.4 \\
$\Psi$ & 1.6 \\
$r$ \tablenotemark{g} & {\bf -2.1} \\
$r_{\rm e}$ \tablenotemark{h} & {\bf -2.3} \\
$n$ & -1.1\\
$b/a$ \tablenotemark{i} & -1.9 \\
$H_{160}$ & 1.4\\
\enddata
\tablenotetext{a}{All values are the number of standard deviations from the null hypothesis that the quantities
are uncorrelated, based on a Spearman rank correlation test.  Positive (negative) values indicate positive (negative) correlation between
the quantities (i.e., larger $r$ corresponds to lower $W_{\lya}$).  The sample size is 71 galaxies.}
\tablenotetext{b}{Stellar mass, estimated from SED fitting.}
\tablenotetext{c}{Average population age from a constant SFR SED model.}
\tablenotetext{d}{Star formation rate measured from SED fitting.}
\tablenotetext{e}{Average star formation rate surface density $\Sigma_{\rm SFR} = SFR/ \pi r_{\rm e}^2$.}
\tablenotetext{f}{Color excess due to dust extinction from a constant SFR SED model.}
\tablenotetext{g}{GALFIT semi-major axis radius.}
\tablenotetext{h}{GALFIT circularized effective radius.}
\tablenotetext{i}{Morphological minor/major axis ratio.}
\label{lyastrength.tab}
\end{deluxetable}

Combining our results with those previously published in the literature, the broad picture of $z \sim 2-3$ \lya\ emitters is therefore relatively clear.
Compared to star forming galaxies  at this epoch with negligible \lya\ emission, galaxies with a measurable \lya\ emission component
tend to be
physically smaller systems with lower stellar mass and less dust, 
lower star formation rates, and a slightly greater degree of nucleation in the rest-frame UV and optical light distribution.

\subsection{Galaxies with Exclusively \lya\ Emission}

Most commonly, galaxies with visible \lya\ emission also show some absorption features in their rest-UV spectra as well
(albeit frequently weak; see discussion by Shapley et al. 2003).  Of the 71 galaxies in the parent sample with \lya\ emission, 7 do not have any identifiable
absorption features (even in a composite stack).  While the \lya\ emission equivalent width for these sources range over a factor $\sim$ 10 ($W_{\lya} = 17 - 184$ \AA),
all seven have Type II morphologies composed of two or more widely separated clumps (see Figure \ref{lyamorphs.fig}), accounting for nearly half of the 17
total galaxies in the parent sample with \lya\ emission and Type II morphology.
By comparison, 100\% of the 28 (26) galaxies with Type I (III) morphology and Ly$\alpha$ emission have visible absorption features in their UV spectra, indicating
that multi-component morphology is a strong factor in determining whether absorption line features will be observed in rest-UV spectra.\footnote{The fraction of galaxies with
Type II morphology and no \lya\ emission that do not have visible rest-UV absorption features is not possible to determine, since by definition such galaxies would
not have spectroscopically confirmed redshifts.}

Morphological indices measuring the two-dimensional distribution of the light from a galaxy (e.g., $A$, $M_{20}$, $\Psi$) therefore easily identify galaxies with
exclusively \lya\ emission in their rest-UV spectra and suggest that these multi-component galaxies are prime major-merger candidates.
If these galaxies are indeed mergers caught in the nearby-pair phase,
the disturbed ISM of the combined system may result in a reduced gas covering fraction, making it easier for resonantly-scattered
\lya\ emission to escape from the system.  However, type II galaxies are no intrinsically brighter in \lya\ emission than Type I or III galaxies; instead, the only significant
difference is that type II galaxies are less likely to have absorption lines in their UV spectra.
A simpler explanation may be that when galaxies with multiple components separated by distances $\sim 1$ arcsec are observed with ground-based spectrographs
the slit tends to be placed midway between the two components, thereby systematically reducing the S/N ratio of the continuum emission against which absorption features
are measured.  As illustrated in Figure \ref{lyamorphs.fig}, in $\sim 50$\% of cases the majority of the rest-UV continuum flux is expected to fall outside the slit
based on the position angles used for the LRIS spectroscopy; even in cases for which the slit was aligned approximately along the separation vector (e.g., Q1623-BX547)
the greater effective area over which the light is distributed may still increase the noise in the continuum such that the observations are systematically biased against
the detection of absorption lines.





%

\section{METAL LINE FEATURES}
\label{metals.sec}

The deepest absorption lines in the rest-UV spectra are the  \sitwo, \oone, \ctwo, \fetwo, and \altwo\ lines that arise in the metal-enriched neutral ISM, which typically
trace massive galactic-scale outflows driven by the intensely starbursting galaxy.  As noted in many
previous publications (e.g., Shapley et al. 2003) there is generally an 
anti-correlation between \lya\ emission and interstellar absorption line strength; since the metal transitions are optically thick declining absorption line strength indicates
a declining gas velocity dispersion and/or covering fraction, which also permits a greater \lya\ escape fraction.

In order to understand the physical mechanisms underlying variations in absorption line strength in greater detail we focus on the HAHQ sample
for which  accurate
systemic redshifts have been obtained from rest-optical nebular emission.
The 35 galaxies in our HAHQ sample represent the subset of the 73 galaxies previously considered by \cite{steidel10} for which {\it HST}/WFC3 rest-optical
imaging data has now been obtained.  In their larger sample, \cite{steidel10}
demonstrated that there is a correlation between baryonic galaxy mass $M_{\rm bary}$ and $\Delta v_{\rm IS}$, the mean velocity offset of the interstellar absorption lines ($z_{\rm IS}$) from the systemic redshift ($z_{H\alpha}$):

\begin{equation}
\Delta v_{\rm IS} = c (z_{\rm IS} - z_{\rm sys})/(1 + z_{\rm sys})
\end{equation}

\begin{deluxetable}{lc}
\tablecolumns{2}
\tablewidth{200pt}
\tabletypesize{\scriptsize}
\tablecaption{Correlations of Bulk Outflow Velocity and Galaxy Properties\tablenotemark{a}}
\tablehead{
\colhead{Quantity} & \colhead{Std. Dev.}}
\startdata
$M_{\rm gas}$\tablenotemark{b} & 0.6\\
$M_{\ast}$\tablenotemark{c} & 1.6\\
$M_{\rm bar}$\tablenotemark{d} & 1.5\\
Age\tablenotemark{e} & 1.9\\
SFR\tablenotemark{f} & -0.9\\
$\Sigma_{\rm SFR}$\tablenotemark{g} & -2.1\\
$E(B-V)$\tablenotemark{h} & 0.7\\
$G$ & -1.7\\
$M_{20}$ & 1.6\\
$C$ & -1.3\\
$A$ & 1.0\\
$\Psi$ & 1.7\\
$r$\tablenotemark{i} & 2.6\\
$r_{\rm e}$\tablenotemark{j} & 2.3\\\
$n$ & -1.6\\
$b/a$\tablenotemark{k} & -0.3\\
$H_{160}$ & 0.3\\
\enddata
\tablenotetext{a}{All values are the number of standard deviations from the null hypothesis that the quantities
are uncorrelated, based on a Spearman rank correlation test.  Positive (negative) values indicate positive (negative) correlation between
the quantities (i.e., larger $r$ corresponds to weaker outflows (less negative $\Delta v_{\rm IS}$).  The sample size is 35 galaxies.}
\tablenotetext{b}{Cold gas mass, estimated from the star formation surface density and the observed H$\alpha$ size, as in Erb et al. (2006c).}
\tablenotetext{c}{Stellar mass, estimated from SED fitting.}
\tablenotetext{d}{Total baryonic mass $M_{\rm bar} = M_{\ast} + M_{\rm gas}$.}
\tablenotetext{e}{Average population age from a constant SFR SED model.}
\tablenotetext{f}{Star formation rate measured from SED fitting.}
\tablenotetext{g}{Average star formation rate surface density $\Sigma_{\rm SFR} = SFR/ \pi r_{\rm e}^2$.}
\tablenotetext{h}{Color excess due to dust extinction from a constant SFR SED model.}
\tablenotetext{i}{GALFIT semi-major axis radius.}
\tablenotetext{j}{GALFIT circularized effective radius.}
\tablenotetext{k}{Morphological minor/major axis ratio.}
\label{voutflow.tab}
\end{deluxetable}

Using the Spearman rank correlation test, \cite{steidel10}  found that the null hypothesis (that $\Delta v_{\rm IS}$ and $M_{\rm bary}$ are uncorrelated) could be ruled out
at $2.7\sigma$ significance.
We repeat the calculations of \cite{steidel10} using an estimate of $M_{\rm bary} = M_{\ast} + M_{\rm gas}$, where $M_{\ast}$ is the stellar mass
derived in Paper I and $M_{\rm gas}$ the gas mass estimated using NIRSPEC and/or OSIRIS H$\alpha$ fluxes in combination with the 
Kennicutt-Schmidt law (\citealt{kennicutt98a}; see discussion by \citealt{erb06c}).
In Table \ref{voutflow.tab} we list the confidence at which the null hypothesis for zero correlation can be rejected for $\Delta v_{\rm IS}$
vs a variety of morphological and other physical properties.
As expected we recover the sense of the previous correlation noted by \cite{steidel10} between \dvis\ and total baryonic mass,
although this correlation has a significance of only $1.5\sigma$; random resampling tests indicate that this lower significance is
consistent with the lower total number of galaxies in our sample than in Steidel et al. (2010).

\subsection{Correlation of Low-ionization Absorption Line Strength with Semi-Major Axis Radius}

We find that \dvis\ correlates with the semi-major axis radius $r$ with a statistical significance of $2.6\sigma$
in the sense that the largest velocity differences (i.e., most negative $\Delta v_{\rm IS}$) are observed for the smallest (Type I) galaxies, while the largest (Type III) galaxies
have mean gas-phase velocities consistent with 0 \kms.
Combining measurements from the individual low-ionization absorption lines (except $\fetwo \, \lambda 1608$, which is significantly weaker and noisier than the other
transitions) we find that $\langle \Delta v_{\rm IS} \rangle = -226 \pm 49$ \kms for the small radius half of the HAHQ sample ($r<1.95$ kpc) 
compared to $\langle \Delta v_{\rm IS} \rangle = -125 \pm 25$ \kms for the large radius half ($r > 1.95$ kpc).\footnote{Uncertainties represent $1\sigma$ uncertainties in the mean
combining $ \Delta v_{\rm IS}$ of individual galaxy spectra.}
We illustrate this trend visually in Figure \ref{LargeRSmallR.fig}; when sorted according to \dvis\ galaxies with the smallest (blue boxes) and largest (red boxes)
sizes are clearly distinguished.

Adopting uncertainties of $\sim 130$ \kms in $\Delta v_{\rm IS}$  and $\sim 0.1$ kpc in $r$ for each galaxy (Steidel et al. 2010; Paper I), we use a least squares routine to fit
a first-order polynomial to the data and obtain the average relation
\begin{equation}
\Delta v_{\rm IS} = (77 \frac{r}{\rm kpc} - 342)\, \kms
\end{equation}
We stress, however, that $\Delta v_{\rm IS}$ does not characterize the outflow velocity of a particular galaxy, but rather represents the centroid of a single-component fit to interstellar absorption
lines that are a complex blend of multiple components.
In Figure \ref{abslines.fig} we show the stacked, continuum-normalized spectra for the small- and large-radius subsamples (solid blue/red lines respectively) in zoomed-in regions
around the major low-ionization gas absorption lines.
Careful inspection of Figure \ref{abslines.fig} suggests that the low ionization absorption lines in both low-$r$ and high-$r$ stacks have similar optical depths in their blue wings, indicating that both have
outflowing gas tracing a similar range of velocities.  Rather, 
the major difference is the growth of an absorption component centered at $v \approx 0$ \kms 
in the large-radius sample that shifts the centroid of  the absorption line profile closer to the systemic velocity (see particularly $\altwo \, \lambda 1671$) and increases 
the depth of composite absorption line
profiles even when precise systemic redshifts are unknown
(e.g., Figure \ref{fullspec.fig}, top panel).
This finding is similar to that described by
Steidel et al. (2010; see also Weiner et al. 2009) who found that galaxies below a threshold $M_{\rm bary} \approx 4 \times 10^{10} M_{\odot}$ lack the $v \sim 0$ \kms component to the low-ionization
interstellar absorption lines present in higher-mass galaxies.

\begin{figure*}
\plotone{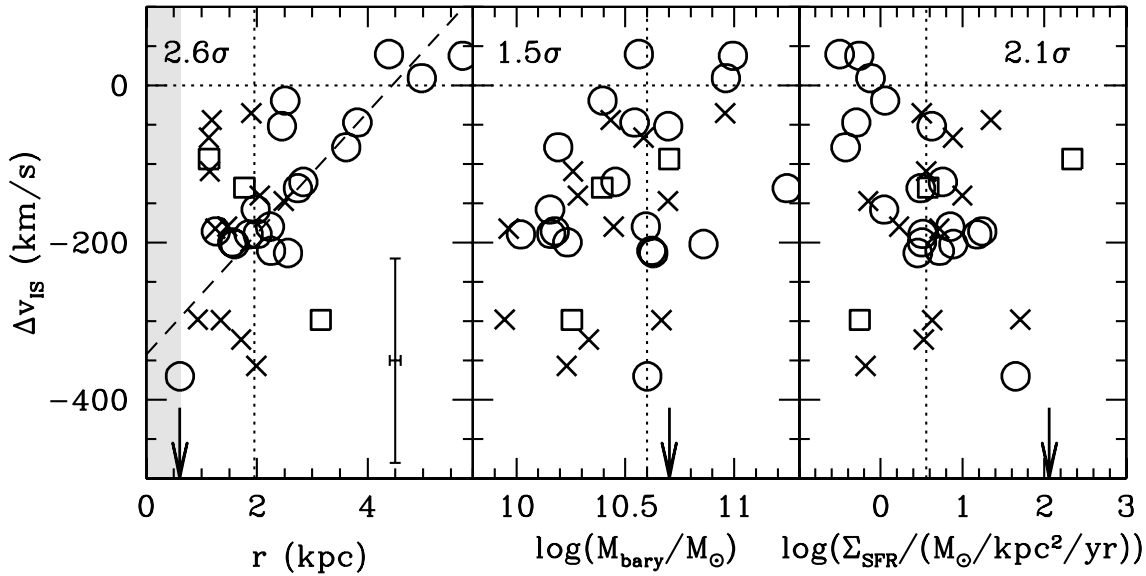}
\caption{Mean rest-frame velocity offset $\Delta v_{\rm IS}$ between the systemic (H$\alpha$) redshift $z_{{\rm H}\alpha}$ and the low-ionization interstellar absorption line redshift $z_{\rm IS}$ 
as a function of major-axis galaxy size $r$ (left panel), baryonic mass (middle panel), and SFR surface density $\Sigma_{\rm SFR}$ (right panel)
for the subset of 35 galaxies with systemic redshifts from H$\alpha$ observations and high quality UV spectra.
The arrow indicates the location of Q2343-BX453 ($\Delta v_{\rm IS} = - 944 \kms$) which along with Q2343-BX587 lies in the region for which $r$ is unresolved at the $3\sigma$ level
(grey shaded region).
Crosses, open boxes, and open circles represent galaxies with morphological Type I, II, and III respectively.
The vertical dotted lines in the left (right) hand panel divide the sample into two equal-size bins of  large-radius and small-radius (large $\Sigma_{\rm SFR}$ and small $\Sigma_{\rm SFR}$) galaxies,
 the vertical dotted line in the middle panel represents
a similar division made according to baryonic mass by Steidel et al. (2010).
Negative/positive values of  $\Delta v_{\rm IS}$ formally correspond to outflows/inflows respectively, note how galaxies with the strongest mean outflow velocities tend to be the smallest
with the largest $\Sigma_{\rm SFR}$
while the three largest galaxies all have positive $\Delta v_{\rm IS}$.
The dashed line indicates a linear least-square fits to the data $\Delta v_{\rm IS} = (77 (r/\textrm{kpc}) \, - 342) \kms$.}
\label{AvelPlot.fig}
\end{figure*}

\begin{figure*}
\plotone{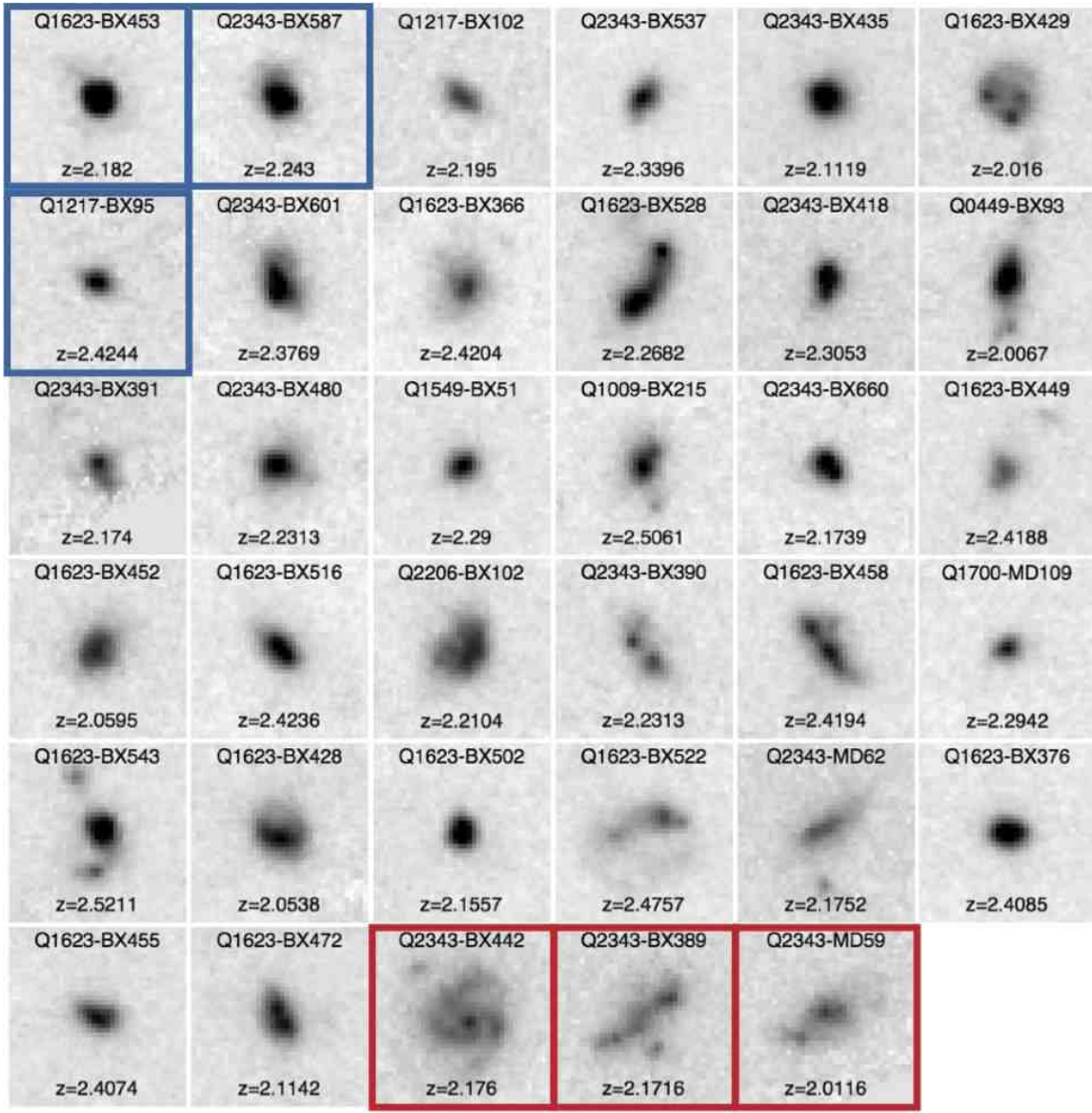}
\caption{{\it HST}/WFC3 F160W morphologies of galaxies in the HAHQ sample (i.e., with H$\alpha$-derived systemic redshifts and high-quality UV spectra) 
sorted (left-right, and top-bottom) in order of declining velocity offset \dvis.
Blue boxes denote the three galaxies with the smallest  semi-major axis radii in the HAHQ sample (Q1623-BX453, Q2343-BX587, Q1217-BX95), red boxes denote
the three galaxies with the largest semi-major axis radii in the sample (Q2343-MD59, Q2343-BX442, Q2343-BX389).
The largest three galaxies all have formal $\dvis > 0$ \kms, while the smallest tend to be those with the most negative $-950 < \dvis < -300$ \kms.
As demonstrated in Figure \ref{fullspec.fig}, the small-$r$ sample has a `younger' UV spectrum with significant Ly$\alpha$ emission.
Individual postage stamps are 3'' to a side, oriented with North up and East left.}
\label{LargeRSmallR.fig}
\end{figure*}

\begin{figure*}
\plotone{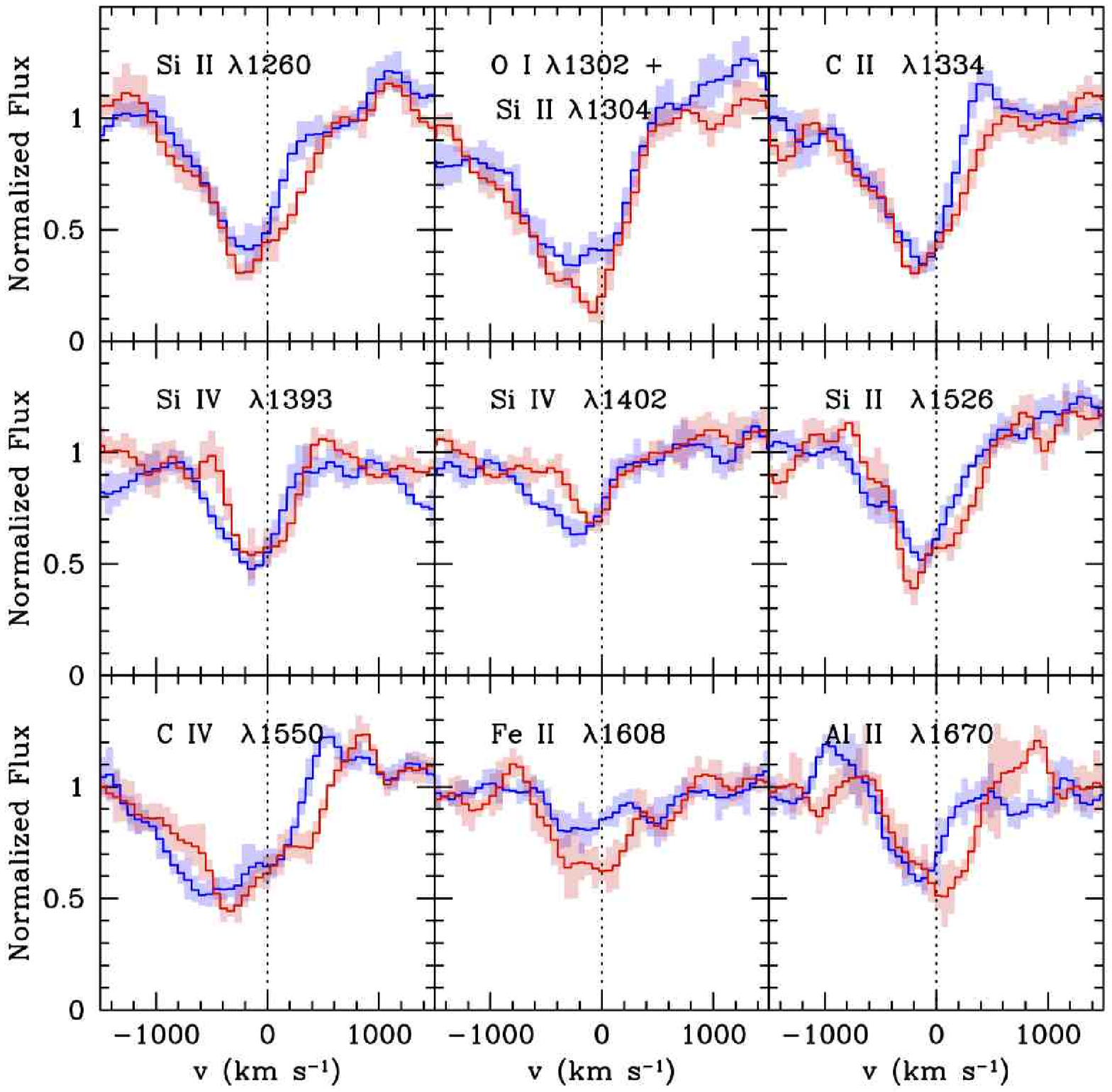}
\caption{As Figure \ref{fullspec.fig} (lower panel) but zoomed in to the central 1500 \kms surrounding each of the major low-ionization
($\sitwo \, \lambda 1260, \oone \, \lambda 1302 $+$ \sitwo \, \lambda 1304, \ctwo \, \lambda 1334, \sitwo \, \lambda 1526, \fetwo \, \lambda1608, \altwo \, \lambda 1670$)
and high-ionization ($\sifour \, \lambda 1393, \sifour \, \lambda 1402, \cfour \, \lambda 1550$) absorption line features.  Blue/red lines respectively represent the stacked
spectra of galaxies in the smallest/largest size bins for the HAHQ sample.  
Shaded regions indicate $1\sigma$ uncertainties estimated by Monte Carlo bootstrap resampling from the individual galaxies
making up the composite spectra.}
\label{abslines.fig}
\end{figure*}

\begin{figure*}
\plotone{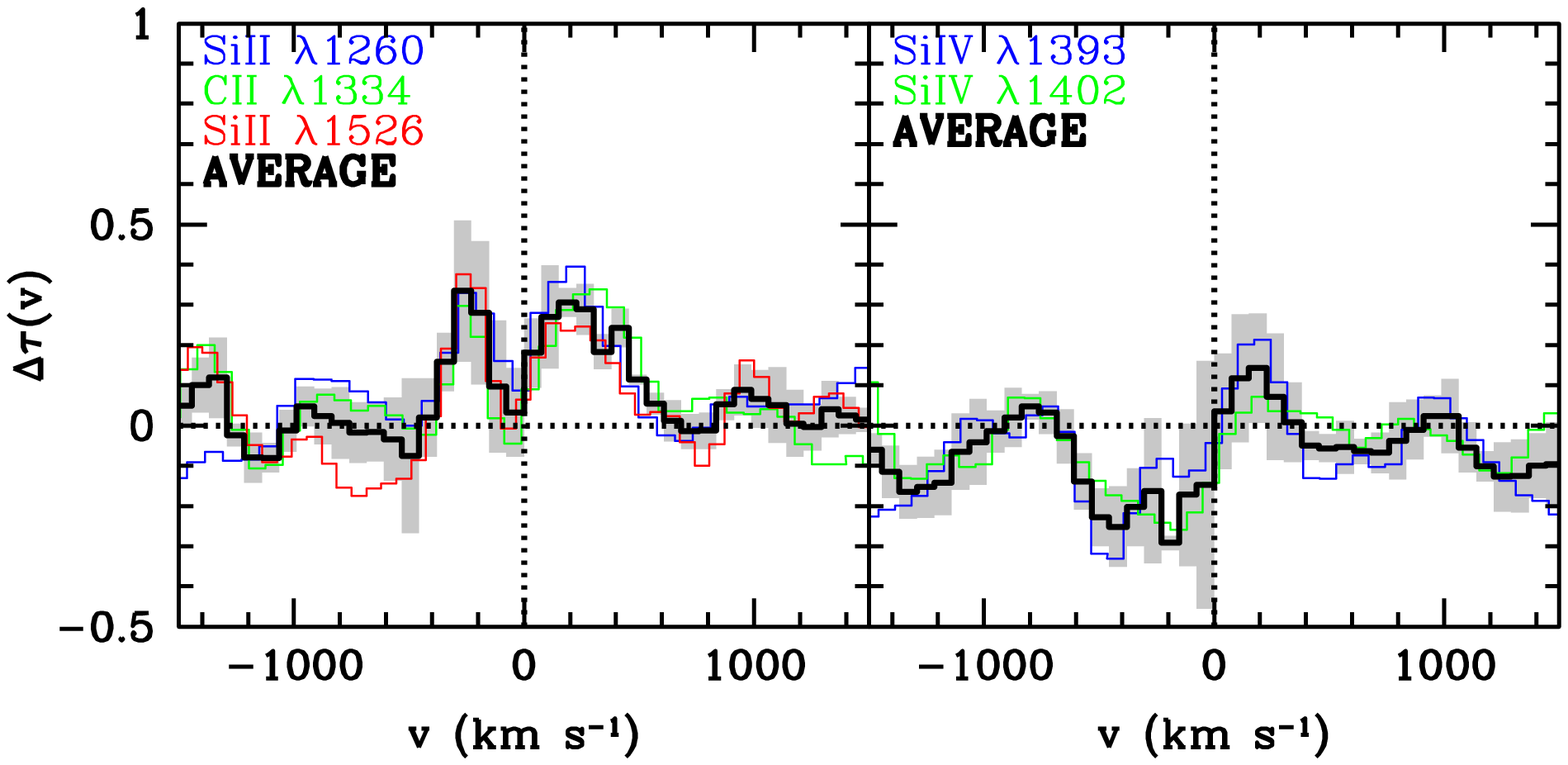}
\caption{Additional optical depth $\Delta \tau(v)$ required to transform the low-$r$ composite galaxy spectrum into the high-$r$ composite as a function of offset from the systemic velocity.
Left-hand panel: Colored lines indicate the profiles derived for individual low-ionization interstellar (LIS) absorption lines, solid black line indicates the average of the three
 individual  $\Delta \tau(v)$ profiles.  Right-hand panel: Colored lines indicate the profiles derived for $\sifour \, \lambda 1393$ and $\lambda 1402$ 
high-ionization interstellar (HIS) absorption lines, solid black line indicates the average of the two individual  $\Delta \tau(v)$ profiles.  Note that the LIS profiles show additional
optical depth near $v \sim 0-200$ \kms\ in large galaxies, while the HIS profiles show marginally 
{\it reduced} optical depth at relative velocities $-600 < v < 0$ \kms.  Shaded grey regions indicate $1\sigma$ uncertainties estimated by Monte Carlo bootstrap
resampling from the individual galaxies
making up the composite spectra.}
\label{tauplot.fig}
\end{figure*}

Following the formalism adopted by Steidel et al. (2010), we define the additional optical depth $\Delta \tau(v)$ that would need to be added to the low-$r$ composite galaxy spectrum in order to produce
an absorption line profile identical to the high-$r$ composite spectrum:
\begin{equation}
I_{\rm hr}(v) = I_{\rm lr}(v) e^{-\Delta \tau(v)}
\end{equation}
where $I_{\rm hr}(v)$ and  $I_{\rm lr}(v)$ are the spectral intensity of the high-$r$ and low-$r$ halves of the HAHQ sample respectively.
In Figure \ref{tauplot.fig} we show $\Delta \tau(v)$ for the three strongest unblended low-ionization absorption lines
($\sitwo \, \lambda 1260$, $\ctwo \, \lambda 1334$, and $\sitwo \, \lambda 1526$)
along with the average composite for all three lines combined.  Although there is some variation between the individual lines, 
the composite exhibits a significant broad absorption component from $\sim -400$ to $\sim +400$ \kms\ in galaxies with the largest semi-major axis radii. 
As discussed at length by Steidel et al. (2010), this  component could represent either infalling gas or stalled winds at large radii, although it most likely
corresponds to cold gas located at small radii within the galaxy.

\subsection{High-ionization absorption lines}

In addition to the low-ionization absorption lines, the high-ionization transitions $\sifour \, \lambda 1393$, $\sifour \, \lambda 1402$, and $\cfour \, \lambda 1550$
are also systematically less blueshifted for larger galaxies.
Combining measurements made from the two $\sifour$ absorption lines,\footnote{We neglect $\cfour$ in this stack since it is complicated by the superposition
 of interstellar absorption on a stellar wind-induced P-Cygni profile.} we find that the mean blueshift of the high ionization lines (HIS) is
$\langle \Delta v_{\rm HIS} \rangle = -147 \pm 58$ \kms for the small-radius sample, but only $\langle \Delta v_{\rm HIS} \rangle = -21 \pm 40$ \kms for the 
large-radius sample.\footnote{Uncertainties represent $1\sigma$ uncertainties in the mean combining $ \Delta v_{\rm HIS}$ of
individual galaxy spectra.}
In contrast to the low-ionization lines however, this effect is not produced by an additional component of $\tau$ centered around $v \sim 0$ \kms,
suggesting that the $v\sim 0$ \kms\ gas may be more strongly self-shielded than the majority of the outflowing gas.
Instead, the large-radius subsample exhibits a  
weaker blue wing to the absorption profile (Figure \ref{tauplot.fig}), suggesting that 
 highly ionized ($T \gtrsim 10^4$ K) gas may have lower peak outflow velocities in galaxies with rest-optical continuum radii $r \gtrsim 3$ kpc.

\subsection{Fine structure emission}
\label{finestruc.sec}

In addition to exhibiting stronger \lya\ emission and more systematically blueshifted low-ionization absorption features,  
galaxies with smaller rest-optical continuum radii also have stronger fine structure emission lines.
As indicated by Figure \ref{fullspec.fig}, $\sitwo^{\ast} \, \lambda 1265$ and $\lambda 1309$ emission features 
 are more prominent
in composite spectra of galaxies with effective radii $r \lesssim 2$ kpc than in those with $r \gtrsim 3$ kpc (most noticeably in the composite stack from the parent sample,
and to a lesser extent in the HAHQ composite stack).
This result is consistent with the recent studies of Berry et al. (2012) and Jones et al. (2012), who
respectively noted that $\sitwo^{\ast}$ increases with $W_{\lya}$ and
is stronger in star forming galaxies at $z\sim4$ than at lower redshifts (which Jones et al. 2012 ascribe to the  increasing characteristic radius of low ionization gas
with cosmic time).

\subsection{Correlation of metal lines with other morphological statistics}
\label{internal.sec}

\begin{figure*}
\plotone{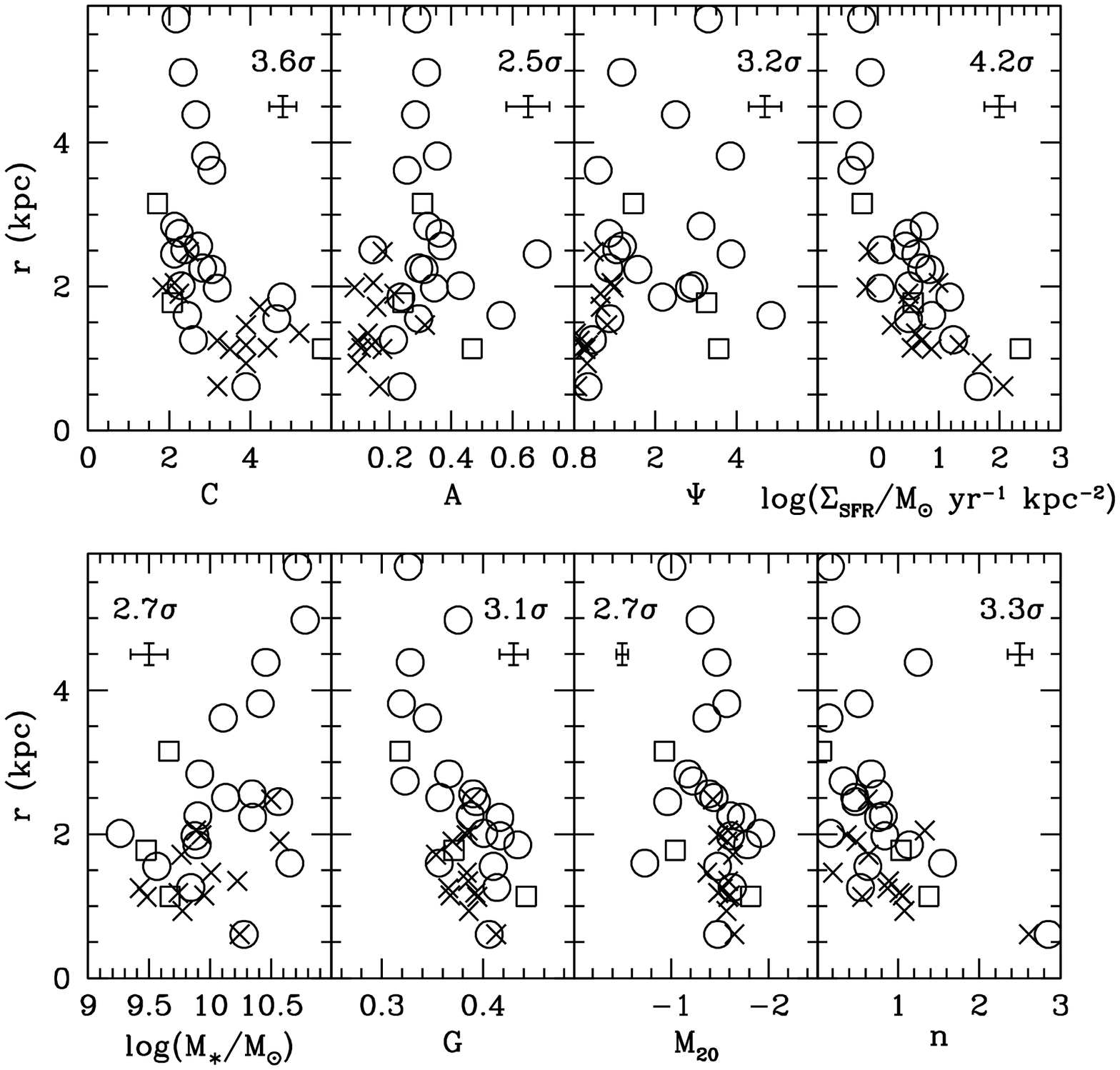}
\caption{Correlations between semi-major axis radius $r$, stellar mass $M_{\ast}$, and the other morphological parameters.  Symbols are as in Figure \ref{AvelPlot.fig}.  Each panel
quotes the statistical significance of the correlation on the basis of the Spearman rank correlation test.  Error bars in
each panel indicate the characteristic uncertainty of individual points.}
\label{internalcorrel.fig}
\end{figure*}

As indicated in Figure \ref{internalcorrel.fig} there are significant underlying correlations between $r$ and the other morphological statistics.
On average, galaxies with larger semi-major axis radii have larger $M_{\ast}$, shallower radial density profiles (lower $n$), less concentrated light profiles (i.e., lower $C$ and $G$),
and tend to be more irregular (whether measured using $M_{20}$, $A$, or $\Psi$).  Such generic trends are apparent in Figure \ref{LargeRSmallR.fig}; 
the largest galaxies tend to be those that show most morphological irregularity to the light profiles, while the smallest tend to be single well-nucleated 
sources.

The physical mechanism underlying the change in \dvis\ is the buildup of neutral gas at rest with respect
to the stars (as discussed further in \S \ref{discussion.sec}).  The internal correlation of semi-major axis radius with other parameters however gives rise to a variety of 
correlations of marginal significance (see Table \ref{voutflow.tab}) between \dvis\ and  stellar/baryonic mass, the six 
morphological statistics $G$, $M_{20}$, $C$,  $A$, $\Psi$, $r$, and $n$, and the SFR surface density $\Sigma_{\rm SFR}$.
Given the known positive correlation between mass and effective radius for the galaxies in our sample (Nagy et al. 2011; Paper I), it is perhaps unsurprising that 
similar qualitative trends in UV spectral features
are observed when the galaxy sample is subdivided according to either mass or radius.
Likewise, although there is no significant correlation of $\Delta v_{\rm IS}$ with the total 
SED-derived star formation rate (in part since all galaxies in our sample are selected on the basis of rest-UV color to 
be actively star forming with SFR
$\gtrsim 2 M_{\odot}$ yr$^{-1}$), there is a correlation
at the $>2\sigma$ level with the SFR surface density in the sense that galaxies with higher $\Sigma_{\rm SFR}$ 
tend to be missing the $v\sim 0$ \kms absorption component  (see Figure \ref{AvelPlot.fig}, right-hand panel).\footnote{Kornei et al. (2012) find that $\Sigma_{\rm SFR}$
is more strongly correlated than galaxy size with \dvis\ for a sample of star forming galaxies at $z\sim1$.}
Similarly, there is a $\sim 2\sigma$ correlation of $\Delta v_{\rm IS}$ with stellar population age in the sense that older populations
have the strongest $v\sim0$ \kms absorption component),
consistent with our finding that the $r<1.95$ and $r>1.95$ kpc samples have mean ages of 260 and 740 Myr respectively.

\section{DISCUSSION}
\label{discussion.sec}

\cite{steidel10} suggested that the physical property governing both \lya\ emission properties and interstellar absorption line kinematics 
was baryonic mass, which correlated with \dvis\ with a significance of
$2.7\sigma$ in their sample of 73 galaxies.  Since
galaxies with higher baryonic masses tend to have older stellar populations and greater total quantities of interstellar gas, this additional gas may naturally lead to greater
covering fractions and longer scattering paths, suppressing \lya\ emission and making interstellar absorption lines deeper.

In detail though, mass may not be the most predictive indicator of the spectroscopic properties of a given galaxy.
Although the correlation that we find between \dvis\ and $r$ is of comparable statistical significance ($2.6\sigma$) to the baryonic mass correlation described by Steidel et al. (2010),
this may simply be because there are $\sim$ a factor of 2 fewer galaxies in the morphological HAHQ sample.  Replicating the HAHQ sample to a total sample size of 73 galaxies
with random resampling using a Monte Carlo algorithm (repeated 10,000 times), 
we estimate that the expected correlation significance between $r$ and \dvis\ in a sample of the size of Steidel et al. (2010) is $3.8\pm0.7 \, \sigma$.
Indeed, while the galaxies with the strongest/weakest \dvis\ occupy an overlapping range of $M_{\rm bary}$ (the galaxy with the {\it weakest} \dvis\ in our
HAHQ sample has a baryonic mass smaller than the galaxy with the {\it strongest} \dvis), these galaxies have clearly distinct visual morphologies
(Figure \ref{LargeRSmallR.fig}).  Likewise, although the galaxies with the highest $\Sigma_{\rm SFR}$ are those with the strongest \dvis\
(consistent with theoretical expectations that galaxies with a higher concentration of supernovae should be capable of more completely expelling the local ISM), this correlation
is driven entirely by the (stronger) relation between $r$ and \dvis\ and the definition $\Sigma_{\rm SFR} = SFR/r^2$ (although c.f. Kornei et al. 2012 for star forming galaxies
at $z\sim1$).

All of the correlations detailed herein may  therefore simply be byproducts of a single underlying physical mechanism; the growth of an extended disk.
Star-forming  $z\sim2-3$ galaxies with low to moderate stellar mass 
(i.e., $M_{\ast} \lesssim 10^{10} M_{\odot}$) tend to be compact, young, triaxial systems whose kinematics are dominated
by velocity dispersion rather than rotation about a preferred kinematic axis.  High $\Sigma_{\rm SFR}$ within these systems
drives strong gaseous outflows into the surrounding IGM and suppresses the buildup of a stable ISM.
In time, the increasing stellar mass formed by these galaxies is able to stabilize the formation of an extended (albeit still thick) gaseous disk 
(as proposed by, e.g., Martig \& Bournaud 2010) in which 
the overall gas fraction decreases ($f_{\rm gas}$ is anticorrelated with \dvis\ with a confidence of $2.8\sigma$ in the HAHQ sample) and
systemic rotation becomes an increasingly
important means of dynamical support.  The lower $\Sigma_{\rm SFR}$ in this extended disk is no longer able to expel gas as efficiently, a large component of which remains close
to the galaxy and serves to both extinguish the Ly$\alpha$ emission
(or resonantly scatter it to larger radii)
 and superimpose a zero-velocity absorption component atop the stellar continuum emission.

While the growth of such extended structure is naturally associated with both evolving $M_{\ast}$ and $\Sigma_{\rm SFR}$, there is not a single well-defined value of either that 
demarcates the transition, and the mass at which extended structures may be supported can vary significantly based on the merger histories and other details of individual galaxies.
In contrast, rest-frame optical size $r$ characterizes where the transition to an extended disk has {\it already occurred}.
From a phenomenological standpoint, rest-optical morphology may therefore be the most robust and accessible means of identifying systems 
for which feedback is less efficient at evacuating gas from the galaxy.



This physical picture necessarily suggests that the zero-velocity gas feature should become more pronounced (i.e., \dvis\ is less strongly blueshifted)
as the contribution of rotational shear to the dynamical support
of a galaxy increases.  Such trends have already been noted in samples of galaxies observed with either long-slit (see Figure 2 of Steidel et al. 2010)
or IFU spectroscopy (see Figure 13 of Law et al. 2009).  More recently, F{\"o}rster Schreiber et al. (2011a) presented {\it HST}/NICMOS morphologies of a sample of six galaxies
observed previously with the SINFONI integral field spectrograph.  Of these six galaxies, five are representative of the most disk-like known systems at $z\sim2$ based on their observed
kinematics (the other galaxy is identified as a major merger).    These five galaxies\footnote{Q1623-BX663, SSA22a-MD41, Q2343-BX389, Q2343-BX610, and Q2346-BX482.  One of these galaxies, Q2343-BX389,
is also in our WFC3 sample.}  all lie at the upper edge of the distribution of effective radii for $z\sim2-3$ star forming galaxies
(see colored circles in Figure 13 of Paper I) and have a mean $\langle \Delta v_{\rm IS} \rangle = -4$ \kms (excluding Q1623-BX663, whose rest-UV spectra 
are complicated by AGN features).
These systems can therefore be identified from the overall $z\sim 2-3$ galaxy sample by both their semi-major axis radii and spatially resolved velocity shear, but not by their stellar masses,
which are widely distributed in the range $M_{\ast} = 8 \times 10^9 - 1 \times 10^{11} M_{\odot}$.

Steidel et al. (2010) suggested that the anticorrelation between \dvis\ and velocity shear may be due in part to projection effects.
If  objects with the largest velocity gradients tend to be edge-on rotating disks, outflows
from these galaxies might be expected to be collimated perpendicular to the disk so that there is simply little projection of the outflowing gas onto our line of sight
(see, e.g., Chen et al. 2010, Bordoloi et al. 2011), while face-on systems would exhibit minimal rotation and maximum \dvis.
Such a correlation between inclination and outflow velocity was recently found for a sample of star forming galaxies at $z\sim1$ by Kornei et al. (2012).
However, in our $z\sim2$ sample we observe no correlation of \dvis\ with inclination as parametrized through the apparent minor/major axis ratio 
$b/a$ for the 35 galaxies in our HAHQ sample ($0.3\sigma$ from the null hypothesis; see Table \ref{voutflow.tab}).
In part, the lack of correlation may simple reflect  the irregularity of $z\sim2$ star forming galaxies and the difficulty of determining inclination robustly
for these intrinsically triaxial (Paper I) systems.  However, no significant difference is observed in \dvis\ even for the two most disk-like galaxies in the sample.
Morphology and IFU-derived kinematics  suggest that Q2343-BX389 is consistent
with an edge-on disk (F{\"o}rster Schreiber et al. 2009), while Q2343-BX442 has both morphology and IFU
kinematics consistent with a nearly face-on ($i = 42^{\circ} \pm 10^{\circ}$) disk with spiral substructure (see discussion by Law et al. 2012b).
Despite their different orientations both galaxies have $\Delta v_{\rm IS} \sim 0$ \kms, suggesting either that \dvis\ is uncorrelated with inclination, or that the
opening angle of the outflow is less than $\sim 40^{\circ}$.

As illustrated by Figure \ref{tauplot.fig} much of the additional absorption in the largest galaxies has positive velocity $\sim 0-400$ \kms with respect to the systemic redshift, and
formally  $\Delta v_{\rm IS} > 0$ \kms\ for the three galaxies with largest semi-major axis radii in our HAHQ sample (Q2343-MD59, Q2343-BX442, Q2343-BX389).
It is therefore possible that absorption line spectroscopy of the largest galaxies is tracing  {\it infalling} gas, whether from a recent merger, recycled wind accretion
(e.g., Oppenheimer et al. 2010), or from cold cosmological flows (e.g., Dekel et al. 2009; Kere{\v s} et al. 2009).
While such inflows are one plausible mechanism for solving the gas fueling probem,
it is unclear why such accretion (presumably with a small covering fraction) would cause the near-complete absorption of Ly$\alpha$ (see
discussion by Steidel et al. 2010).
Similarly,  simulations (e.g., Dekel \& Birnboim 2006; Kere{\v s} et al. 2009) generally predict that such cold flows should be suppressed
in the most massive galaxies at $z>2$, in contrast to our observation that $\Delta v_{\rm IS} > 0$ \kms\ occurs more commonly for large and massive galaxies.

\section{SUMMARY}
\label{summary.sec}

We have demonstrated that rest-optical morphology is correlated with the gas-phase properties of $z\sim2$ 
star forming galaxies as traced by their rest-UV spectra.  \lya\ emission is most commonly observed in
galaxies with small rest-optical half-light radii ($r < 2$ kpc), and the equivalent width $W_{\lya}$ of the emission is 
correlated with half light radius with $> 2\sigma$ confidence.
Although some large-$r$ galaxies with Type III morphology (i.e., extended and diffuse) show \lya\ in emission, 
this emission is relatively weak; the twelve galaxies in our sample with $W_{\lya} > 30$ \AA\ all have small half-light radii ($r < 2$ kpc)
and Type I or Type II morphologies (i.e., compact clumps).
This correlation is independent of the better-known relations between $W_{\lya}$, dust
content, and star formation rate.  Additionally, all galaxies in our spectroscopic sample for which \lya\ is observed in emission and no absorption lines
are present in the rest-UV spectrum have morphological Type II (i.e., consist of two or more clumps of emission separated 
from each other by $\sim$ 1 arcsec), suggesting that slit losses of continuum light from these galaxies may be significant.

Combining our results with those previously published in the literature (e.g., \citealt{steidel10}, \citealt{penterrici10}, \citealt{kornei10}), 
the broad picture of $z \sim 2-3$ star forming galaxies is relatively clear.
Galaxies with \lya\ emission tend to be physically smaller than galaxies with negligible  \lya\ emission,
with lower stellar mass, less dust, lower star formation rates, and a slightly greater degree of nucleation in the rest-frame UV and optical light distribution.

Using a subsample of 35 galaxies with both high-quality rest-UV spectra and H$\alpha$-derived systemic redshifts
we have found that galaxy morphology is also correlated with the median blueshift of interstellar absorption line features \dvis\ with $2.6\sigma$ significance.
Although similar blue wings in the low-ionization absorption line profiles indicate that galaxies of all morphological types
drive comparably strong outflows at speeds up to $\sim 800$ \kms\ into their surrounding environments,
increasing optical half-light radius is accompanied by an increase in optical depth around the systemic redshift of the galaxy.
This finding is similar in statistical significance to trends with increasing baryonic mass described recently by \cite{steidel10}, but may be of physically greater
importance considering the smaller sample size.
While galaxies with  \dvis\ differing by almost 1000 \kms\ can have baryonic masses  identical to within observational uncertainty, 
such galaxies have clearly distinct morphologies
spanning the range from compact clumps to extended disks.

The increased optical depth arises in gas nearly at rest with respect to the systemic redshift, in close physical proximity to the galaxy, 
and apparently corresponding to an extended disk of optical continuum light. 
This stronger $v\sim0$ \kms\ component is responsible both for the shifting absorption line profile
and the attenuation of $W_{\lya}$ (via a longer resonant scattering path) in galaxies with larger rest-optical half light radii.
The $v\sim0$ \kms\ absorption component is not present in the higher-ionization features $\sifour \, \lambda 1393$ and $1402$ however, suggesting that
this gas may be more strongly self-shielded than the majority of the outflowing gas.

In contrast to recent  results at lower redshifts (e.g., \citealt{kornei12}), there is no obvious relation between \dvis\ and
galaxy inclination as parameterized by the axis ratio $b/a$ for the 35 galaxies in our HAHQ sample.  
In part this may be due to the difficulty determining robust inclinations for inherently triaxial
star-forming systems.  Even for two particularly well-resolved galaxies with visual morphology and IFU kinematics consistent with face-on and edge-on disks
however, there are no obvious differences in the outflowing gas kinematics (although the quality of individual spectra prohibits a detailed analysis of just
these two galaxies).  There is therefore no evidence to support a classical bipolar outflow model for star forming galaxies at $z\sim2-3$.  This may be because
such galaxies are in the process of transforming from clumpy, irregular systems  to more regular (albeit thick) disks, and outflowing gas at
high velocities and large distances from the galaxy may predate establishment of the disk and its corresponding polar axis.


In general, our observations are consistent with inside-out growth of star forming galaxies in the young universe:
`Typical' $z\sim2$ star forming galaxies appear to be gas-rich,
compact, triaxial systems systems that are dominated by velocity dispersion between individual star forming regions rather than systemic rotation, and whose high $\Sigma_{\rm SFR}$
drives strong outflows into the surrounding IGM.  As these galaxies mature they gain stellar mass, stabilizing the formation of extended (albeit still thick) gaseous disks
in which rotational support is observed to play an increasing role.  As the star formation migrates from central regions 
into these extended disks 
the $\Sigma_{\rm SFR}$ drops and is no longer capable of expelling gas from the galaxy as efficiently.

We caution, however, that it is challenging to generalize about the relation between rest-optical morphology and gas phase kinematics 
for the entire $z\sim2$ star forming sample
on the basis of the relatively small number of galaxies in our sample.  
A major limiting factor is systemic nebular redshifts, without which we cannot quantify the outflow velocities.
Although $\sim 100$ $z\sim2-3$ star forming  galaxies with rest-UV spectra in the KBSS have had systemic redshifts derived from nebular emission line
spectroscopy (e.g., Erb et al. 2006b; Law et al. 2009), only a fraction of these galaxies fell within the footprint of our {\it HST}/WFC3 imaging survey.
In the future it will be possible to explore these trends in much greater detail with the aid of multi-object near-IR and optical spectrographs, using which it will be possible to obtain
H$\alpha$ redshifts and outflow kinematics efficiently for large and diverse samples of galaxies in fields that have already been imaged with {\it HST}.

\acknowledgements

DRL, CCS, and SRN have been supported by grant GO-11694 from the Space Telescope Science Institute,
which is operated by the Association of Universities for Research in Astronomy, Inc., for NASA, under contract NAS 5-26555.
CCS has been supported by the US National Science Foundation through grants AST-0606912 and AST-0908805
AES acknowledges support from the David and Lucile Packard Foundation.
Finally, we extend thanks to those of Hawaiian ancestry on whose sacred mountain we are privileged to be guests.

\end{document}